\documentclass[prb,superscriptaddress,showpacs,aps]{revtex4}
\usepackage{graphicx}
\usepackage{dcolumn}
\usepackage{bm}
\usepackage{amsmath}
\usepackage{color}
\usepackage{epsfig}

\usepackage[T2A]{fontenc}
\usepackage[cp1251]{inputenc}

\begin{document}


\title{Gyrokinetic theory of magnetic structures in high-$\beta$ plasmas of the Earth's magnetopause and of the slow solar wind}

\author{Du\v san Jovanovi\'c}
\email{dusan.jovanovic@ipb.ac.rs} \affiliation{Institute of Physics, University of Belgrade, Pregrevica 118, 11080 Belgrade (Zemun), Serbia}
\author{Olga Alexandrova}
\email{olga.alexandrova@obspm.fr} \affiliation{Observatoire de Paris--Meudon, Laboratoire d'Etudes Spatiales et d'Instrumentation en Astrophysique (LESIA), Centre National de la Recherche Scientifique (CNRS), Meudon, France}
\author{Milan Maksimovi\'c}
\email{milan.maksimovic@obspm.fr} \affiliation{Observatoire de Paris--Meudon, Laboratoire d'Etudes Spatiales et d'Instrumentation en Astrophysique (LESIA), Centre National de la Recherche Scientifique (CNRS), Meudon, France}
\author{Milivoj Beli\'c}
\email{milivoj.belic@qatar.tamu.edu} \affiliation{Texas A\&M University at Qatar, P.O. Box 23874 Doha, Qatar}
\date{\today}

\begin{abstract}
Nonlinear effects of the trapping of resonant particles by the combined action of the electric field and the magnetic mirror force is studied using a gyrokinetic description that includes the finite Larmor radius effects. A general nonlinear solution is found that is supported by the nonlinearity arising from the resonant particles, trapped by the combined action of the parallel electric field and the magnetic mirror force. Applying these results to the space plasma conditions, we demonstrate that in the magnetosheath plasma, coherent nonlinear magnetic depression may be created associated with the nonlinear mirror mode and supported by the population of trapped ions forming a hump in the distribution function. These objects may appear either isolated or as the train of weakly correlated structures (the cnoidal wave). In the Solar wind and in the Earth's magnetopause, characterized with anisotropic electron and ion temperatures that are of the same order of magnitude, we find coherent magnetic holes of the same form that are attributed to the two branches of the nonlinear magnetosonic mode, the electron mirror and the field swelling mode, including also the kinetic Alfv\'{e}n mode, and supported by the population of trapped electrons. The localized magnetic holes may have the form of a moving oblique slab or of an ellipsoid parallel to the magnetic field and strongly elongated along it, that propagates along the magnetic field and may be convected in the perpendicular direction by a plasma flow. While the ion mirror structures are purely compressional magnetic, featuring negligible magnetic torsion and electric field, the magnetosonic and kinetic Alfv\'{e}n structures possess a finite electrostatic potential, magnetic compression, and magnetic torsion, but the ratio of the perpendicular and parallel magnetic fields remains small.

\end{abstract}

\maketitle

\section{Introduction}\label{Introductory}

Coherent magnetic structures are ubiquitous in the space plasma of the solar system, where they have been observed over the full range of distances and latitudes relative to the Sun. They were detected in the solar
wind \cite{Denises_paper,7,7a}, in the Earth's magnetosphere, i.e. in the magnetotail \cite{4} and in the magnetopause \cite{10,1a,2017NatCo...814719G}, in the magnetospheres of the Mars, Saturn, and Jupiter \cite{14,2,9}, and also in the induced magnetospheres of Venus, Io, and comets \cite{5,3,6,15}. The Voyager mission detected magnetic structures in the heliosheath, beyond the heliospheric termination shock \cite{8}. Magnetic structures mostly have the form of solitary magnetic depressions (holes) or the trains of magnetic holes \cite{1a}. Solitary magnetic humps were detected less frequently, e.g. in the Earth and Jovian magnetosheaths \cite{11,12}, while trains of humps and the combinations of humps and holes were observed in the Earth's magnetosheath \cite{13}. 
Magnetic structures often feature a large perturbation of the intensity of the magnetic field (10--50\%, sometimes \cite{1a} as large as 98\%), and very little bending. They are pressure balanced, i.e. exhibit the anticorrelation between the magnetic and thermal pressures. Their perpendicular scale is several, to several tens proton gyroradii, but holes of several hundreds gyroradii have also also detected \cite{1a}. Their pitch angle to the magnetic field is close to $90^{\rm o}$, yielding the aspect ratio of 7--10.

In the sheath plasmas the thermal pressure exceeds the magnetic pressure and we often have $\beta > 10$, where $\beta = 2 p/c^2 \epsilon_0 B^2$ is the ratio of thermal and magnetic pressures. The ion temperature is usually both anisotropic and much larger than the electron temperature, $T_{i_\bot} > T_{i_\Vert} > T_e$. Under such conditions, several linear modes are unstable. The thermal anisotropy in a high-$\beta$ plasma drives both the ion mirror mode \cite{16,19}, whose parallel phase speed is much smaller than the ion thermal speed, and the ion cyclotron mode \cite{18,19}. In the same range of phase speeds, the halo in the tail of the distribution function drives the halo instability \cite{Pokhotelov_FLR_2}. Conversely, in the magnetopause the electron and ion temperatures are close to each other and the temperature anisotropy is usually not very large, but there exist a strong current that yields the magnetic reconnection, contributing also to the creation of magnetic structures \cite{1a}. Moreover, the coexistent inhomogeneity of the pressure and magnetic field, via the Hall instability, destabilizes the kinetic Alfv\'en wave \cite{2005ChA&A..29....1D} that propagates faster than the ion acoustic speed and slower than the electron thermal speed. The gradient and the anisotropy of the electron temperature, for certain combinations of the plasma $\beta$ and of the anisotropy $T_{e_\bot}/T_{e\Vert}$, can also excite the instabilities \cite{1982PhRvL..48..799B} of the magnetosonic mode, 
whose parallel phase velocity lies between the parallel electron and ion thermal speeds. In the literature, the unstable fast magnetosonic mode is referred to as the \textit{field swelling mode} and the unstable slow magnetosonic mode as the \textit{electron mirror instability}, for details see e.g. \cite{17}. Particularly important is the short wavelength limit of the ideal MHD fast magnetosonic mode, in which the spatial scale of disturbances is close to the ion Larmor radius or to the electron inertial length, and the Alfv\'{e}n mode acquires an electric field parallel to the background magnetic field. Such mode is usually referred to as the {\em kinetic Alfv\'{e}n} mode. It often occurs in space physics where it is responsible for the acceleration and energization of particles as well as for the exchange of energy between waves and particles. Linear kinetic Alfv\'{e}n waves are unstable in the presence of inhomogeneities of the density and magnetic field \cite{2005ChA&A..29....1D} and of the electron temperature anisotropy \cite{2010PhPl...17f2107C}.

Because of such richness of linear instabilities, that presumably saturate into the magnetic structures \cite{1a,14,2,9,5,3,6,15,4,10,2017NatCo...814719G,7,7a,8,11,12,13}, 
the nature of the latter still remains elusive. The prevailing theory relates them with the nonlinear mirror mode, but other models have also been proposed, based on the magnetic reconnection \cite{24},  magnetohydrodynamic (MHD) beam microinstabilities \cite{22}, Hall MHD of charge-exchange processes \cite{8}, and on magnetosonic solitons \cite{1a,10,20}.

The linear \textit{mirror mode} in a spatially uniform, bi-Maxwellian ($T_\Vert\ne T_\bot$) plasma is weakly dispersive due to the finite ion Larmor radius effects. Under magnetosheath conditions, when the electrons are cold and massless, the mirror mode is purely growing \cite{19} 
due to the resonant contribution of the particles with the zero parallel velocity. However, 1-D particle simulations \cite{26,29} revealed that the saturation of the mirror instability produced humps rather than holes, if the linear drive was strong enough. In weakly unstable configurations periodic structures with moderate humps and holes were obtained, while under linearly stable conditions initially imposed holes persisted for very long times. Accordingly, mirror-mode humps are observed in the middle of the magnetosheath, while the holes are observed close to the magnetopause \cite{27}, where the mirror mode is marginally stable. In most cases, the trains of humps are created rather than isolated humps, and on a very long timescale these are inverted to become holes \cite{25}. The saturation mechanism, for a strong drive, comes from the trapping of the resonant ions by the mirror force, producing vortices in the phase space, which actually dominates the mirror mode dynamics in the case of a weak drive \cite{25,26}.

KdV-type \textit{magnetosonic solitons} exist in the case of propagation at sufficiently large angles to the magnetic field \cite{1969JPSJ...27.1331K,1986PhFl...29.1844O}. Conversely, for a quasiparallel propagation, envelope solitons become possible \cite{20} that are essentially Alfv\'{e}n wave packets modulated by zero-frequency acoustic perturbations and described by the derivative nonlinear Schr\"{o}dinger equation (DNSE). However, both the KdV and the DNSE equations describe the dynamics of finite (but small!) amplitude perturbations of the compressional magnetic field that are strictly 1-D (slab) structures, unstable in the transverse direction. The soliton theory has been criticized \cite{8}, because it requires a quasiparallel propagation, in a sharp disagreement with the observed large aspect ratios. Conversely, 1-D slow magnetosonic solitons propagate at close to $90^{\rm o}$ to the magnetic field \cite{10}. On the proton scale there may also exist \textit{electrostatically charged magnetic structures}, whose self-organization comes from the nonlinear effects associated with trapped electrons and the magnetic hole or a hump is created by the current of the $\vec{E}\times\vec{B}$ drift of trapped electrons \cite{2012AnGeo..30..711T}. Such magnetized electron phase-space holes have been observed in the plasma sheath \cite{2015JGRA..120.2600S} and in 2-D PIC simulations \cite{2015PhPl...22a2309H}. Moreover, phase-space structures can be driven also by the grad-B and $\vec{E}\times\vec{B}$ currents of ions that are trapped in a self-consistent magnetic bottle \cite{2009PhPl...16h2901J}.

\textit{Magnetic bubbles} were observed by the Polar satellite \cite{1a} in the high-latitude magnetopause boundary and in the presence of strong magnetopause currents (i.e. near a possible reconnection site). They featured strong depressions (up to 98\%) of the ambient magnetic field and were filled with heated solar wind plasma and immersed in a broadband turbulent spectrum of kinetic Alfv\'{e}n waves. Numerical simulations \cite{1a} indicated that the bubbles could be produced by the magnetic reconnection, with the accompanying kinetic Alfv\'{e}n fluctuations coming from the Hall instability driven by the macroscopic gradients of pressure and magnetic field. A similar situation has been recently revisited by the NASA's Magnetospheric Multiscale (MMS) mission \cite{2017NatCo...814719G} that enabled 3-d measurements of both the charged particles and the electromagnetic fields, with a sufficiently high resolution to resolve the ion kinetic scale (i.e. the scale of the ion Larmor radius). The MMS mission observed compressive fluctuations featuring anti-correlated perturbations of the electron density and the magnetic field magnitude, in the vicinity of a recent magnetic reconnection that produced a plasma jet flowing nearly anti-parallel to the background magnetic field with a speed $\sim c_S\sim 0.5\; c_A$, where $c_S$ and $c_A$ are the acoustic and the Alfv\'{e}n speeds, respectively. These magnetic field fluctuations and bursts of electron phase space holes
appeared together with the kinetic Alfv\'{e}n wave in the locations of strong electron pressure gradients. The magnetic structure had the form of a kinetic Alfv\'{e}n wave packet, propagating at the pitch angle $\sim 100^{\rm o}$ to the ambient magnetic field, that exhibited spatial structure in the transverse direction, of the order of an ion gyroradius. The close examination of the electron velocity distribution function in the wave packet revealed that besides the isotropic thermal core and two suprathermal beams counterstreaming along the magnetic field, commonly observed in the magnetopause boundary layer, there existed also a population of trapped particles which accounted for $\sim 50\%$ of the density fluctuations and a $\sim 20\%$ increase in the electron temperature within the KAW. However, the latter was not indicative of heating but rather of a nonlinear capture process that may have provided the nonlinear saturation of Landau and transit-time damping. These electrons were trapped within adjacent wave peaks by the combined effects of the parallel electric field and the magnetic mirror force. Their distribution function unmistakably exhibited the loss-cone features, since it contained only the particle velocities with near $90^{\rm o}$ magnetic pitch angles. In the magnetic hole recorded by \cite{2017NatCo...814719G}, the ratio of the minimum to maximum magnetic field magnitude was $B_{min}/B_{max}\sim 0.96$, and the resulting magnetic mirror force was sufficient to trap electrons with magnetic pitch angles between $75^{\rm o}$ and $105^{\rm o}$.

In the present paper, we study the effects of particle trapping in a high-$\beta$ plasma with anisotropic temperature, using the Chew--Goldber--Law gyrokinetic theory and including the Dippolito--Davidson treatment of  higher-order corrections \cite{1966PhFl....9.1475F,1967PhFl...10..669D,1975PhFl...18.1507D}. We derive the nonlinear equation for the compressional magnetic field, 
allowing also for a finite parallel electric field (the latter is short-circuited only when the electrons are cold), including also the convection of both particle species by the grad-B drift. In the stationary regime, the appropriate expressions for the energy, magnetic moment, and canonical momentum for both species are found,
and used to construct their distribution functions \cite{Luque-Schamel}. In appropriate limits, our equations reduce to the nonlinear ion mirror \cite{19,Kuznetsov_JETPL,2009PhPl...16h2901J}, kinetic Alfv\'{e}n \cite{2017NatCo...814719G}, electron mirror-, and the field swelling modes \cite{17}, as well as to the magnetized electrostatic electron and ion holes \cite{2012AnGeo..30..711T}. We demonstrate that in the general case, all perturbations whose characteristic perpendicular scale exceeds the ion scales (i.e. the ion plasma length, the ion Larmor radius, or the ion acoustic radius) are described by the same generic nonlinear equation (\ref{eq_for_zeta}), which possesses two distinct coherent solutions in the form of a slab that is oblique to the magnetic field and propagates perpendicularly to it, or of a finite length filament ('cigar') parallel to the magnetic and propagating along the latter. A propagating, infinitely long, oblique filament, i.e. a cylinder with ellipsoidal cross section, is also possible but its description requires the solution of a 2-D nonlinear equation and it has been left out from our present study. Our oblique slab is, actually, the limiting case of the well known periodic {\em cnoidal wave} solution \cite{Luque-Schamel} that can fully reproduce the properties of the Ref.  \cite{2017NatCo...814719G} structures. Conversely, our filaments are fundamentally different from the high-$\beta$ MHD (quasi)monopolar vortices, governed by the fluid convective nonlinearity, which are prohibited in the kinetic Alfv\'{e}n regime  \cite{Rad_za_Olgu_i_Milana}.

\section{Gyrokinetic description of perturbations in a warm plasma, somewhat bigger than the ion-scale}

In order to study the effects of the mirror force on plasma particles, we use the classical Chew--Goldber--Law gyrokinetic theory, including the Dippolito--Davidson treatment of  higher-order corrections \cite{1966PhFl....9.1475F,1967PhFl...10..669D,1975PhFl...18.1507D}.
The latter is obtained by the integration of the Vlasov equation for the particles' gyroangle, taking that the dynamics of the particles' guiding centers is slow on the temporal scale of the their cyclotron gyrations, that the dynamics of magnetic field lines belongs to the same slow temporal scale and that their curvature is relatively small. 
It includes the terms of the zeroth and of the first order in the small parameter
pertinent to the drift scaling and to the small corrections coming from the finite Larmor radius and from the displacement current, viz.
\begin{equation}\label{gyro_drift_scaling}
\omega/\Omega \sim \omega/\omega_{p e}\sim k_\bot\rho_L \sim \epsilon \ll 1,
\end{equation}
where $\omega$ and $k_\bot$ are the characteristic frequency and characteristic perpendicular wavenumber. The Dippolito--Davidson theory was developed under the ordering
$v_{T_\bot} B  \sim \vec{E}_\bot \sim \epsilon^{-1} E_\Vert $ and $k_\Vert\sim k_\bot, $ which resulted in a rather complicated gyrokinetic equation (1) of Ref. \cite{1967PhFl...10..669D}. The latter is considerably simplified if we relax their ordering between the parallel and perpendicular wavenumbers as well as for parallel and perpendicular electric fields. Here, in addition to the constraints Eq. (\ref{gyro_drift_scaling}), we assume a weak $z$-dependence, an electric field that is mostly perpendicular to the magnetic fiel, and small perturbations of the density and of the magnetic field, viz.
\begin{equation}\label{My_scaling}
\omega/\Omega \sim \omega/\omega_{p e}\sim k_\Vert/ k_\bot\sim \epsilon \quad {\rm and} \quad \epsilon \; v_{T_\Vert} B \lesssim \epsilon \; v_{T_\bot} B \sim \vec{E}_\bot  \sim \epsilon^{-1} E_\Vert ,
\end{equation}
and the gyrokinetic equation of Refs. \cite{1966PhFl....9.1475F,1967PhFl...10..669D,1975PhFl...18.1507D} obtains an elegant form
{
that is accurate to the leading order in $\epsilon$, viz.
\begin{equation}\label{my_gyrokinetic_2}
\left[\frac{\partial}{\partial t} + v_\Vert\vec{b}\cdot\nabla + \left(\vec{V}_E + \vec{V}_B\right)\cdot\nabla_\bot + a_\Vert \frac{\partial}{\partial v_\Vert} + a_\bot \frac{\partial}{\partial v_\bot}\right] f\left(\vec{r}, v_\Vert, v_\bot\right) = 0,
\end{equation}
}where $\vec{v}$ is the particle velocity, while $v_\Vert = \vec{b}\cdot\vec{v}$ and $v_\bot = |\vec{b}\times\vec{v}|$ are the magnitudes of its components parallel and perpendicular to the magnetic field, respectively. The guiding-centers' distribution function $f(\vec{r}, v_\Vert, v_\bot)$ is obtained by the integration of the particle distribution function $f(\vec{r}, \vec{v})$ for the gyroangle $\theta$, defined as $\theta = \arccos(\vec{v}\cdot\vec{n})$, where $\vec{n}$ is a unit vector in the direction of the bi-normal of the magnetic field line, $\vec{n} = \vec{b}\times (\nabla\times\vec{b})/ |\nabla\times\vec{b}|$. Here $\vec{V}_E$ is the $\vec{E}\times\vec{B}$ drift velocity, while $\vec{V}_p$, $\vec{V}_B$, and $V_\Vert$ are the kinetic counterparts of the grad-B, polarization, and parallel drift velocities. The parallel acceleration $a_\Vert$ comes from the electric field and from the mirror force, while the perpendicular acceleration $a_\bot$ is equal to the divergence of the guiding center velocity, viz.
\begin{equation}\label{kinetic_drifts}
\vec{V}_E = -\frac{\vec{b}}{B}\times\vec{E}, \quad
\vec{V}_B = \frac{v_\bot^2}{2\Omega}\frac{\vec{b}}{B}\times\nabla_\bot B, \quad
\vec{V}_p = \frac{\vec{b}}{\Omega}\times\left[\frac{\partial}{\partial t} + \left(v_\Vert\vec{b} + \vec{V}_E\right)\cdot\nabla\right]\left(v_\Vert\vec{b} + \vec{V}_E\right), \quad
V_\Vert = \frac{v_\bot^2}{2\Omega}\frac{\vec{b}}{B}\,\cdot\,\left(\nabla\times\vec{B}\right),
\end{equation}
\begin{equation}\label{kinetic_accels}
a_\Vert = \vec{b}\cdot\left\{\frac{q}{m}\;\vec{E} - \frac{v_\bot^2}{2 B}\; \nabla B - \frac{v_\bot^2}{2 \Omega}\; \nabla\times\left[\frac{\partial}{\partial t} + \left(v_\Vert\vec{b} + \vec{V}_E\right)\cdot\nabla\right] \vec{b}\right\}, \quad
a_\bot = -2 v_\bot\nabla_\bot\cdot\left(\vec{V}_E + \vec{V}_B + \vec{V}_p + v_\Vert \vec{b} \; \right).
\end{equation}
Here $\Omega$ is the gyrofrequency, $\Omega= q B/m$.
Velocities and accelerations given in Eqs. (\ref{my_gyrokinetic_2})-(\ref{kinetic_accels}) have been calculated with the accuracy to second order in the small parameter $\epsilon$ introduced in the ordering of Eq.
(\ref{My_scaling}), viz.
\begin{equation}\label{small_param}
\frac{1}{\Omega}\frac{\partial}{\partial t} \sim \frac{1}{\Omega} \, \left(\vec{V}_E + \vec{V}_B\right)\cdot\nabla  \sim \frac{\vec{b} \, \cdot\nabla}{\nabla_\bot}
\sim \frac{|\vec{E}_\Vert|}{|\vec{E}_\bot|} \sim\frac{\delta n}{n} \sim \frac{|\delta\vec{B}|}{|\vec{B}|}\sim\frac{|\vec{V}_E + \vec{V}_B|}{v_{T_\Vert}}
\sim\epsilon \ll 1.
\end{equation}
In Ref. \cite{2009PhPl...16h2901J}, a gyrokinetic equation has been derived that permits also large perturbations of the compressional magnetic field, if the curvature of the magnetic field lines is sufficiently small so that magnetic curvature and helicity can be neglected in the (small) terms coming from the ion polarization by grad-B drift. In other words, when the unperturbed magnetic field is oriented along the $z$-axis, viz. $\vec{B}_0 = \vec{e}_z B_0$, the results of Ref. \cite{2009PhPl...16h2901J} are applicable when $\delta B_z\sim B_z$, but $|\vec{B}|/B_z -1 \sim\epsilon^2$. Note that the scaling of Eq. (\ref{small_param})
does not set a strong constraint on the Larmor radius, since  it gives
\begin{equation}\label{dodatno_skaliranje}
\frac{|\vec{V}_E|}{v_{T_\Vert}}\sim \frac{v_{T_\bot}}{v_{T_\Vert}} \, \frac{|v_{T_\bot}\nabla_\bot|}{\Omega} \,\frac{q\phi}{T_\bot}\sim\epsilon
\quad \Rightarrow \quad
\frac{|\vec{V}_B|}{|\vec{V}_E|}\sim\frac{\delta B_z/B_0}{q\phi/T_\bot}\sim\frac{|v_{T_\bot}\nabla_\bot|}{\Omega}\equiv\delta,
\end{equation}
where $\delta$ appears to be of arbitrary order.
However, although it is not self-evident, our gyrokinetic equation may be valid when $\delta = |v_{T_\bot} \nabla_\bot|/\Omega \sim \epsilon^\frac{1}{2}$. To demonstrate this, we deduce from Eq. (\ref{my_gyrokinetic_2}) the corresponding hydrodynamic equations of continuity and parallel momentum, and compare them with
the hydrodynamic equations that exist in the literature, whose domain of validity and accuracy with respect to the small parameters $\epsilon$ and $\delta$ are known.
Integrating the gyrokinetic equation (\ref{my_gyrokinetic_2}) in velocity space and with appropriate weight functions $1$ and $v_\Vert$ and after some tedious but straightforward algebra, carefully keeping the leading terms in the small parameter $\epsilon$, we arrive at
\begin{eqnarray}
\nonumber
&&
\left[\frac{\partial}{\partial t} + \left(\vec{U}_E+\vec{U}_D\right)\cdot\nabla_\bot\right]\left(\log n - \log B\right) + \left(\vec{b}\cdot\nabla\right)\left[U_\Vert + \frac{p_{\Vert_0} - p_{\bot_0}}{q n_0 B_0} \;\; \nabla\cdot\left(\vec{e}_z\times\vec{b}\right)\right] + \\
&& \frac{1}{\Omega_0}\nabla_\bot\cdot\left[\left(\frac{\partial}{\partial t} + \vec{U}_E\cdot\nabla_\bot\right)\vec{e}_z\times\vec{U}_E\right] = 0,
\label{new_continuity_gyro}
\\
&&
\label{parallel_momentum_aprox_gyro}
\left[\frac{\partial}{\partial t} + \left(\vec{U}_E + \vec{U}_B\right)\cdot\nabla\right]U_\Vert =
\frac{q}{m} \; \vec{b}\cdot\vec{E} - \frac{1}{n_0 m}\left(\vec{b}\cdot\nabla\right)\left[p_\Vert - \left(p_{\Vert_0} - p_{\bot_0}\right)\log{B}
\right],
\end{eqnarray}
for the notations, see Eq.s (\ref{drift_velocity_1}) and (\ref{drifts}).
Equations (\ref{new_continuity_gyro}) and (\ref{parallel_momentum_aprox_gyro}) include the effects of particles' gyromotion, through the convection by the grad-$B$ drift and the acceleration by the mirror force, which makes them more general than the standard Strauss's equations of reduced MHD  \cite{1976PhFl...19..134S,1977PhFl...20.1354S} in a moderately cold plasma, $\beta_{i_\bot} \sim \beta_{e_\bot}\lesssim 1$, 
from which the mirror force is absent.
{More accurate fluid calculations (see e.g. Ref. \cite{2015PhyS...90h8002J}) include also finite Larmor radius corrections to the convective derivative $\vec{U}_E\cdot\nabla$, and the diamagnetic and grad-B contributions to the plasma polarization, viz.
\begin{eqnarray}
\nonumber
&& \hspace{-1cm}
\left(\frac{\partial}{\partial t} + \vec{U}_\bot^{apr}\cdot\nabla_\bot\right)\left(\log n - \log B\right) + \left(\vec{b}\cdot\nabla\right)\left[U_\Vert + \frac{p_{\Vert_0} - p_{\bot_0}}{q n_0 B_0} \;\; \nabla\cdot\left(\vec{e}_z\times\vec{b}\right)\right] + \\
&& \hspace{-1cm}
\frac{1}{\Omega_0}\nabla_\bot\cdot\left\{\left[\frac{\partial}{\partial t} + \left(\vec{U}_\bot^{apr} + \vec{U}_B - \vec{U}_D\right)\cdot\nabla_\bot\right]\vec{e}_z\times\vec{U}_\bot^{apr}\right\} = 0,
\label{new_continuity}
\\
&&
\hspace{-1cm}
\label{parallel_momentum_aprox}
\left[\frac{\partial}{\partial t} + \left(\vec{U}_\bot^{apr} + \vec{U}_B - \vec{U}_D\right)\cdot\nabla\right]U_\Vert =
\frac{q}{m} \; \vec{b}\cdot\vec{E} - \frac{1}{n_0 m}\left(\vec{b}\cdot\nabla\right)\left[p_\Vert - \left(p_{\Vert_0} - p_{\bot_0}\right)\log{B} - \frac{p_{\bot_0}}{\Omega_0} \; \nabla_\bot\cdot\left(\vec{e}_z \times \vec{U}_\bot^{apr}\right)\right],
\end{eqnarray}
where $\vec{U}_\bot^{apr} = (1 - v_{T_\bot}^2\nabla^2_\bot/2\Omega^2)^{-1}(\vec{U}_E + \vec{U}_D)$.}
Obviously, our moment equations (\ref{new_continuity_gyro}) and (\ref{parallel_momentum_aprox_gyro}) agree with the more accurate fluid equations (\ref{new_continuity}) and (\ref{parallel_momentum_aprox})
in the regime of small Larmor radius corrections, $
\rho_{Li}\nabla_\bot \sim \epsilon^\frac{1}{2}$, when both the density perturbations are sufficiently small $\delta n/n\ll q\phi/T_\bot$ and the nonlinear convection by grad-B drift in the polarization term can be neglected. The latter is possible not only when $\delta B/B\ll q\phi/T_\bot$, that is realized in low-$\beta$ plasmas, but also for $\delta B/B\sim q\phi/T_\bot$ in 1-D slab and cylindrically symmetric geometries, in which the essentially 1-D shape of the structure suppresses all convective derivatives.
In view of this, we conclude that the gyrokinetic equation (\ref{my_gyrokinetic_2}) can be used with caution also in plasmas with large ratios of thermodynamic and magnetic pressures, $\beta_\bot>1$, to describe kinetic phenomena whose characteristic scales are somewhat bigger than the Larmor radius, $\rho_{Li}\nabla_\bot < 1$.

\subsection{Integrals of motion (characteristics of the gyrokinetic equation)}

We take that the unperturbed magnetic field is oriented along the $z$-axis, viz. $\vec{B}_0 = \vec{e}_z B_0$, and seek a localized, stationary, 2-D solution of Eq. (\ref{my_gyrokinetic_2}) that is travelling with the velocity $\vec{e}_z \, u_z$, where $u_z$ is an arbitrary phase velocity. This implies that the solution depends only on the variables $v_\Vert$, $v_\bot$, $x$, $y$, and $z_1 = z - t/u_z$. Then, using $\partial/\partial t= \partial/\partial t'
-u_z \, \partial/\partial z$, the gyrokinetic equation (\ref{my_gyrokinetic_2}) can be rewritten as
\begin{equation}\label{my_gyrokinetic_2_travel-tilt}
\left[\frac{\partial}{\partial t'} +
\left(v_\Vert-u_z\right)\frac{\partial}{\partial z} + \left(\vec{e}_z \times\nabla_\bot \chi\right)\cdot\nabla_\bot + a_\Vert \frac{\partial}{\partial v_\Vert} + a_\bot \frac{\partial}{\partial v_\bot}\right] f\left(x, y, z_1, v_\Vert, v_\bot\right) = 0,
\end{equation}
where, for a stationary solution, we set $\partial/\partial t'= 0$ and keeping only the terms of the orders $\epsilon^2$ and $\epsilon^2\delta$, we also have
\begin{eqnarray}
\label{chi}&&
\chi = \frac{1}{B_0}\left(\phi - v_\Vert A_z +\frac{m v_\bot^2}{2 q}\,\log{B}\right),
\\
\label{a_par}&&
a_\Vert = -\frac{1}{v_\Vert-u_z}\left[\left(v_\Vert-u_z\right)\frac{\partial}{\partial z} + \left(\vec{e_z}\times\nabla_\bot\chi\right)\cdot\nabla\right] \left[\frac{q}{m}\left(\phi-u_z A_z\right) + \frac{v_\bot^2}{2}\;\log B\right],
\\
\label{a_nor}&&
a_\bot = \frac{v_\bot}{2}\left[\left(v_\Vert-u_z\right)\frac{\partial}{\partial z} + \left(\vec{e_z}\times\nabla_\bot\chi\right)\cdot\nabla\right]\left[\log{B} + \frac{1}{\Omega B_0}\,\nabla_\bot^2\left(\phi - v_\Vert A_z\right)\right].
\end{eqnarray}
The characteristics of the above stationary, 3-d gyrokinetic equation are determined from
\begin{equation}\label{characteristics}
\frac{-d x}{\partial\chi/\partial y_1} = \frac{d y
}{\partial\chi/\partial x} = \frac{d z_1}{v_\Vert-u_z} = \frac{d v_\Vert}{a_\Vert} = \frac{d v_\bot}{a_\bot},
\end{equation}
from which we can calculate explicitly only two integrals of motion, the energy $W$ and the magnetic moment $\mu$, viz.
\begin{eqnarray}\label{ener}
&& W = \left({m}/{2}\right)\left[v_\bot^2 + \left(v_\Vert-u_z\right)^2\right] + q\left(\phi-u_zA_z\right),
\\\label{mag_mom}
&& \mu = \log{\frac{B}{v_\bot^2}} + \frac{1}{\Omega B_0}\nabla_\bot^2\left(\phi-v_\Vert A_z\right).
\end{eqnarray}
Our expression (\ref{mag_mom}) for the magnetic moment coincides with that derived by Davidson \cite{1967PhFl...10..669D} within the less restrictive gyro-drift scaling of Eq. (\ref{gyro_drift_scaling}), and also (in the appropriate limit) with the result of Jovanovi\'c and Shukla \cite{2009PhPl...16h2901J} that permits also large perturbations of the compressional magnetic field.
In a special case of a 2-D solution that is tilted relative to the $z$ axis by the small angle $\theta\approx\tan\theta = u_y/u_z\ll 1$ (where $u_y$ is the $y$-component of the phase velocity), for which we have $\partial/\partial y = (u_z/u_y)\;\partial/\partial z$, we find one more conserved quantity, identified as the canonical momentum $P$, viz.
\begin{equation}\label{mom}
P = m\left(v_\Vert-u_z\right) + q\left(A_z -x B_0 u_y/u_z\right).
\end{equation}
Such tilted solution depends on four variables, $v_\Vert$, $v_\bot$, $x$, and $y_1 = y +(u_y/u_z)\; z - t/u_y$, and the conserved quantities (\ref{ener})-(\ref{mom}) constitute a complete set. Thus, an arbitrary travelling-tilted 2-D distribution function can be expressed as the function of three variables $W$, $\mu$, and $P$. As the last one contains the explicit spatial variable $x$, a distribution function can feature a $P$-dependence only if it is spatially dependent in the unperturbed state.
It should be noted also that the above integrals of motion have been calculated with the accuracy to first order in the small parameter $\epsilon$, where in the expressions for the energy and the canonical momentum we neglected small terms of order ${\cal O}(\epsilon \, v_\bot^2\nabla^2/\Omega^2)$, while the small variation of the magnetic moment is given by
$
d\mu = \left({1}/{B_0^2}\right)\nabla_\bot^2 A_z \; dv_\Vert.
$ 
In a strictly 1-D case, this gives $d\mu = (1/2 B_0^2) \; d \, \{[\nabla_\bot(A_z-x B_0 u_y/u_z)]^2\} \sim \epsilon^2 d$, and we expect that $d\mu$ is of the same order also in 2-D and  3-D.

\subsection{Free and trapped particles}\label{FRT}

The stationary state under study has been established at a distant past $t\to-\infty$ and thus the solution of the 2-D stationary gyrokinetic equation (\ref{my_gyrokinetic_2}) is constant along its characteristics. From the conservation laws (\ref{ener}) and (\ref{mag_mom}) we can relate the particle velocities at the infinity, $v_\bot^{(0)}$ and $v_\Vert^{(0)}$, with those at the phase-space location $(x, y_1, v_\bot, v_\Vert)$. These "initial velocities" are the functions of integrals of motion and within the adopted accuracy take the form
\begin{eqnarray}
\label{init_vel_nor}
&&
v_\bot^{(0)} = \sqrt{B_0 \, \exp(-\mu)} = v_\bot \left\{1 - \frac{1}{2} \left[\frac{\delta B}{B_0} + \frac{1}{\Omega B_0}\nabla_\bot^2\left(\phi - v_\Vert A_z\right)\right]\right\} + {\cal O}\left(\epsilon^2\right)
\\
\label{init_vel_par}
&&
v_\Vert^{(0)} = u_z \pm \sqrt{{2 W}/{m} - B_0 \, \exp\left(-\mu\right)} = u_z + \sigma \sqrt{\left(v_\Vert - u_z - \delta u_z\right)^2 - \Delta^2 } + {\cal O}\left(\epsilon^2\right)
\end{eqnarray}
where
\begin{equation}\label{def_dve_delte}
\delta u_z = \frac{v_\bot^2\nabla_\bot^2}{2 \Omega B_0} A_z,
\quad\quad
\Delta = \sqrt{- 2\frac{q}{m}\left(1+\frac{v_\bot^2\nabla_\bot^2}{2 \Omega^2}\right)\left(\phi - u_z A_z\right) - v_\bot^2 \frac{\delta B}{B_0} },
\quad\quad \textrm{and} \quad\quad
\sigma = {\rm sign}\left(v_\Vert - u_z - \delta u_z\right).
\end{equation}
One should keep in mind that cold and massless electrons efficiently short-circuit the parallel electric field and that, as a consequence, the term $(q/m)(\phi - u_z A_z)$ may become very small. As the latter appears to be the leading term within the scalings (\ref{small_param}) and (\ref{dodatno_skaliranje}), it is necessary that we retain also the next-order term $(v_\bot^2 \nabla_\bot^2 /2 \Omega B_0) u_z A_z$ in Eqs. (\ref{init_vel_nor}) and (\ref{def_dve_delte}) although, at first sight, it appears to be a small quantity of higher order.

We take that that the electromagnetic field is localized, i.e. that that the potentials $\phi$ and $A_z$, and the compressional field $\delta B_z$ vanish at infinity, $\phi, A_z, \delta B_z\to 0$ when $r\to\infty$, where $r=\sqrt{x^2+y_1^2}$. In such a case, there exist two fundamentally different shapes of the characteristics, i.e. of the particle trajectories in phase space $(x, y_1, v_\bot, v_\Vert)$, determined by $W = {\rm constant}, \mu = {\rm constant}, P = {\rm constant}$:
\\
\textit{i)} Open characteristics, stretching to an infinitely distant point in real space $r\to\infty$. Particles following open characteristics are labeled as \textit{free}.
\\
\textit{ii)} Characteristics that close on themselves and are confined to a limited domain in phase space. Particles on such trajectories are \textit{trapped}.

On open characteristics, the distribution function is equal to its asymptotic value at $r\to\infty$, which we adopt to be a Maxwellian with anisotropic temperature, viz.
\begin{equation}\label{free_distr}
f^{free}\left(v_\Vert, v_\bot\right) = f_0\left(v_\Vert^{(0)}, v_\bot^{(0)}\right) = \frac{n_0}{\sqrt{2\pi} \; v_{T_\Vert}\, v_{T_\bot}^2} \; \; \exp\left(-\frac{m{v_\Vert^{(0)}}^2}{2 T_\Vert}-\frac{m{v_\bot^{(0)}}^2}{2 T_\bot}\right).
\end{equation}
Here $n_0$ is the unperturbed ion density, $T_\bot$ and $T_\Vert$ are the perpendicular and the parallel (to the magnetic field) ion temperatures, respectively, and $v_{T_\bot}$ and $v_{T_\Vert}$ are the corresponding thermal velocities, $v_{T_{\bot,\Vert}}=\sqrt{T_{\bot,\Vert}/m}$. The "initial velocities" $v_\Vert^{(0)}$ and $v_\bot^{(0)}$ are given in Eqs. (\ref{init_vel_par}) and (\ref{init_vel_nor}).
Clearly, the initial parallel velocity $v_\Vert^{(0)}$ of free particles 
must be a real quantity, which is realized when $2 W/m \geq B_0 \, \exp(-\mu)$.
In the simple case $\phi - u_z A_z=0$, yielding $\Delta = v_\bot \sqrt{-\delta B/B_0}$, we find that inside a local minimum of the magnetic field, $\delta B<0$, the velocities of free particles belong to the \textit{loss cone} in velocity space $(v_\Vert - u_z)/v_\bot > \sqrt{-\delta B/B_0}$.

Conversely, for $2 W/m < B_0 \, \exp(-\mu)$, i.e. for particles whose parallel velocities are in the region $v_\Vert\in(u_z + \delta u_z - \Delta,$  $u_z + \delta u_z + \Delta)$, the corresponding "initial velocity" $v_\Vert^{(0)}$ is a complex quantity. Such result is unphysical and it implies that these particles have never been at, and will never come to, an asymptotic location $r\to\infty$. In other words, these particles are trapped on their characteristics which are closed curves in phase-space. As the particle trajectories do not cross, such closed characteristics occupy a region in phase-space that is inaccessible for free particles. This further implies that, in a distant past, the trapped particles have gone through some nonadiabatic process, during which time the potentials and the compressional magnetic field have been time-dependent, the term $\partial f/\partial t'$ in the gyrokinetic equation (\ref{my_gyrokinetic_2_travel-tilt}) has been finite and their energy and magnetic moment have not been conserved. Over time, trapped particles perform a large number of bounces and we expect that the phase-averaging of the individual trajectories of trapped particles results in a shifted thermal distribution, with a parallel temperature $T_\Vert^{trap}$, viz.
\begin{equation}\label{trap_distr}
f^{trap}\left(v_\Vert, v_\bot\right) = \frac{n_0\;\; \exp\left(-m u_z^2/2 T_\Vert\right)}{\sqrt{2\pi} \; v_{T_\Vert}\, v_{T_\bot}^2}
 \; \; \exp\left[-\frac{m}{2 T_\Vert^{trap}}\left({v_\Vert^{(0)}}-u_z\right)^2 -\frac{m{v_\bot^{(0)}}^2}{2 T_\bot}\right],
\end{equation}
where such normalization has been adopted that the distribution functions $f^{free}$ and $f^{trap}$ are continuous at the branch point $v_\Vert^{(0)}=u_z$, determined by Eq. (\ref{init_vel_par}). It should be noted that the trapped particles are isolated from those that are free and that the parallel temperature of trapped particles $T_\Vert^{trap}$ may be different than that of free particles and it can be even negative. This does not contradict the second law of thermodynamics, since trapped particles occupy only a limited phase-space volume within which their distribution function remains finite, irrespectively of the sign of the temperature. As a consequence, the relevant integrals of distribution function also remain finite.

Now we can calculate the necessary hydrodynamic quantities as the moments of the particle distribution function, performing the integration in velocity space with appropriate weight functions. It is instructive to separate nonresonant and resonant contributions in a specific moment $\cal M$, denoted by the superscripts "$nr$" and "$res$", as follows
\begin{eqnarray}\label{moments_nr}
&&{\cal M}^{nr} = \int_0^\infty v_\bot \, dv_\bot \;\; p\int_{-\infty}^\infty dv_\Vert \; \xi\left(v_\Vert, v_\bot\right) \, f^{free},
\\
&&
\label{moments_res}
{\cal M}^{res} = \int_0^\infty v_\bot \, dv_\bot \;\; p\int_{u_z+\delta u_z - \Delta}^{u_z+\delta u_z + \Delta} dv_\Vert \; \xi\left(v_\Vert, v_\bot\right) \, f^{res},
\end{eqnarray}
where $p$ denotes the principal value of an integral. In the above, the resonant distribution function $f^{res}$ is defined as $f^{res} = f^{trap}-f^{free}$ and the weight function $\xi$ takes the values $\xi = 1$, $v_\Vert$, $m v_\Vert^2$, and $m v_\bot^2/2$, in the expressions for the number density $n$, the parallel hydrodynamic flow $n \, U_\Vert$, the parallel pressure $p_\Vert$, and the perpendicular pressure $p_\bot$, respectively.
In the computation of nonresonant contributions, Eq. (\ref{moments_nr}), we conveniently expand the free distribution function $f^{free}$ using the small quantity $\Delta^2/(v_\Vert - u_z - \delta u_z)^2$, which permits us to rewrite $f^{free}$ in the form
\begin{equation}
f^{free} = \frac{n_0  \; \exp\left(-\frac{m v_\Vert^2}{2 T_\Vert}-\frac{m v_\bot^2}{2 T_\bot}\right)}{\sqrt{2\pi} \; v_{T_\Vert}\, v_{T_\bot}^2} \;
\left[1+\frac{m}{2 T_\Vert} \frac{v_\Vert}{v_\Vert-u_z} \, \Delta^2 +\frac{m v_\bot^2}{2 T_\bot}\left(\frac{\delta B}{B_0} + \frac{\nabla_\bot^2\phi}{\Omega B_0}\right)+ v_\Vert \left(\frac{1}{T_\Vert}-\frac{1}{T_\bot} \right) \frac{m v_\bot^2}{2\Omega B_0} \; \nabla_\bot^2 A_z\right].
\label{free_distr_appr}
\end{equation}
This enables a straightforward integration in Eq. (\ref{moments_nr}), yielding
\begin{eqnarray}
\label{n_nr}
&&
\frac{n^{nr}}{n_0} = 1-Z_R\left[\left(1+ \frac{v_{T_\bot}^2\nabla_\bot^2}{\Omega^2}\right)\frac{q}{T_\Vert}\left(\phi-u_z A_z\right) + \frac{T_\bot}{T_\Vert}\frac{\delta B}{B_0}\right] + \frac{\delta B}{B_0} + \frac{\nabla_\bot^2\phi}{\Omega B_0},
\\
\label{Up_nr}
&&
\frac{U_\Vert^{nr}}{u_z}
= - Z_R\left[\left(1+ \frac{v_{T_\bot}^2\nabla_\bot^2}{\Omega^2}\right)\frac{q}{T_\Vert}\left(\phi-u_z A_z\right)+\frac{T_\bot}{T_\Vert}\frac{\delta B}{B_0}\right] - \frac{v_{T_\Vert}^2}{u_z^2}\left(1 - \frac{T_\bot}{T_\Vert}\right)\frac{\nabla_\bot^2 A_z}{\Omega B_0},
\\
\label{pp_nr}
&&
\frac{p_\Vert^{nr}}{n_0 T_\Vert} = 1 - \left(1 + \frac{u_z^2}{v_{T_\Vert}^2} \; Z_R\right)\left[\left(1 + \frac{v_{T_\bot}^2\nabla_\bot^2}{\Omega^2}\right) \frac{q}{T_\Vert}\left(\phi-u_z A_z\right) + \frac{T_\bot}{T_\Vert}\frac{\delta B}{B_0}\right] + \frac{\delta B}{B_0} + \frac{\nabla_\bot^2\phi}{\Omega B_0},
\\
\label{pn_nr}
&&
\frac{p_\bot^{nr}}{n_0 T_\bot} = 1 - Z_R\left[\left(1 + 2\frac{v_{T_\bot}^2\nabla_\bot^2}{\Omega^2}\right) \frac{q}{T_\Vert}\left(\phi-u_z A_z\right) + 2\frac{T_\bot}{T_\Vert}\frac{\delta B}{B_0}\right] + 2\frac{\delta B}{B_0} + 2\frac{\nabla_\bot^2\phi}{\Omega B_0},
\end{eqnarray}
The parameter $Z_R$ is the real part of the Fried-Conte plasma dispersion function $Z_\Vert$ (also called the $z$-function):
\begin{equation}\label{plasma_disp}
Z_\Vert = Z_R + i \, Z_I \equiv \frac{1}{\sqrt{2\pi} v_{T_\Vert}}
\left[p\int_{-\infty}^\infty \frac{dv_\Vert \; v_\Vert}{v_\Vert-u_z}\;\exp\left(-\frac{m v_\Vert^2}{2T_\Vert}\right) \, + i \pi \, u_z \, \exp\left(-\frac{m u_z^2}{2T_\Vert}\right)\right],
\end{equation}
and it has simple asymptotic values $Z_R\to 1$ for $u_z\ll v_{T_\Vert}$ and $Z_R\to -v_{T\Vert}^2/u_z^2$ for $u_z\gg v_{T_\Vert}$.
{Although the finite Larmor radius terms in Eqs. (\ref{n_nr})-(\ref{pn_nr})
are small within our scaling, viz. $(v_{T_\bot}^2\nabla_\bot^2/\Omega^2)(\phi-u_z A_z)\to 0$, we may need them later since they provide the dispersion of MHD-like modes, such as the field swelling, electron-, and ion-mirror modes. As already mentioned, FLR terms are not calculated accurately from the gyrokinetic \cite{1966PhFl....9.1475F,1967PhFl...10..669D,1975PhFl...18.1507D} equation (\ref{my_gyrokinetic_2}). Making a comparison with the solutions of fluid equations (\ref{new_continuity}) and (\ref{parallel_momentum_aprox}) [see also Eq.s (16) and (17) in Ref. \cite{2015PhyS...90h8002J}], we note that an appropriate description of the grad-$B$, polarization, and weak FLR effects is obtained when in Eqs. (\ref{n_nr})-(\ref{pn_nr})
we implement the substitution $\nabla_\bot^2\phi\to \nabla_\bot^2(\phi + p_\bot/q n_0)$, where the leading-order expression for $p_\bot$ is used,
$p_\bot\approx n_0 T_\bot[1-Z_R \,\; (q/T_\Vert)(\phi-u_z A_z) + 2(1-Z_R \,\; T_\bot/T_\Vert)(\delta B/B_0)]$.
We note also that the densities and parallel fluid velocities of nonresonant particles, Eqs. (\ref{n_nr})-(\ref{pn_nr}),
have the same form as the linear solutions of the fluid equations (\ref{new_continuity_gyro})-(\ref{parallel_momentum_aprox_gyro}). This implies that the adopted simple form (\ref{free_distr}) of the distribution function exist only when the sought-for coherent nonlinear structure possesses a geometry for which the nonlinearities due to convective derivatives vanish (e.g. 1-D slab or cylindrically symmetric geometries).}

In Eq. (\ref{moments_res}), the integration is performed over the domain of trapped particles, viz. $v_\Vert-u_z\lesssim\Delta\sim\epsilon^\frac{1}{2}$, inside  which it is convenient to rewrite $f^{free}$ in the following way
\begin{equation}\label{f_free_res}
f^{free}\left(v_\Vert, v_\bot\right) = \frac{n_0\;\; \exp\left(-\frac{m u_z^2}{2 T_\Vert}\right)}{\sqrt{2\pi} \; v_{T_\Vert}\, v_{T_\bot}^2}
 \; \; \exp\left[-\frac{m}{2 T_\Vert}\left({v_\Vert^{(0)}}-u_z\right)^2 -\frac{m{v_\bot^{(0)}}^2}{2 T_\bot}\right] \exp\left[-\frac{m u_z}{T_\Vert}\left(v_\Vert^{(0)} - u_z\right)\right],
\end{equation}
which is further simplified setting
\begin{equation}\label{exp_aprox}
\exp\left[-\frac{m u_z}{T_\Vert}\left(v_\Vert^{(0)} - u_z\right)\right]\approx \exp\left[\frac{m^2 u_z^2}{2 T_\Vert^2}\left(v_\Vert^{(0)} - u_z\right)^2\right]-\frac{m u_z}{T_\Vert}\left(v_\Vert^{(0)} - u_z\right).
\end{equation}
Using the above and comparing Eq.s (\ref{f_free_res}) and (\ref{trap_distr}), we note that in the domain of resonant parallel velocities, the trapped particles' distribution  $f^{trap}$ and the even part of $f^{free}$ have identical forms, but with different parallel temperatures. Now we can easily write down the effective distribution in the resonant domain, viz.
\begin{eqnarray}
\nonumber
&&
\hspace{-1cm}
f^{res} = \frac{n_0}{\sqrt{2\pi} \; v_{T_\Vert}\, v_{T_\bot}^2} \; \; \exp\left(-\frac{m u_z^2}{2 T_\Vert}-\frac{m v_\bot^2}{2 T_\bot}\right) \; \frac{m}{2 T_\Vert}\times
\\
&&
\hspace{-1cm}
\left\{\left(1-\frac{m u_z^2}{T_\Vert} - \frac{T_\Vert}{T_\Vert^{trap}}\right)\left[\left(v_\Vert - u_z-\delta u_z\right)^2 - \Delta^2\right] +
\frac{u_z}{v_\Vert-u_z-\delta u_z} \left[2 \left(v_\Vert - u_z-\delta u_z\right)^2 -\Delta^2\right]\right\},
\label{f^res}
\end{eqnarray}
where small terms of order $\leq\epsilon^{3/2}$ have been neglected and we have conveniently separated even and odd parts.

Finally, performing the integrations in velocity space, we obtain the moments Eq. (\ref{moments_res}) in a closed form, as
\begin{eqnarray}
\label{n_res_konacno}
&&
\frac{n^{res}}{n_0} = - R_\Vert\left(\frac{T_\bot}{T_\Vert}\right)^\frac{3}{2}\left[1 + \frac{3}{2}\frac{\zeta_1}{\varphi} +
\frac{3}{4}\frac{\zeta_1^2}{\varphi^2} \,\; g\left(\varphi/\zeta_1\right)\right]\; \varphi^\frac{3}{2} ,
\\
\label{U_res_konacno}
&&
\frac{\left(n \, U_\Vert\right)^{res}}{n_0 \, u_z} = \left(1 + Q_\Vert\right)
\;\frac{n^{res}}{n_0},
\\
&&
\label{pp_res_konacno}
\frac{p_\Vert^{res}}{n_0 T_\Vert} = \frac{u_z^2}{v_{T_\Vert}^2}
\left(1 + 2 Q_\Vert\right)
\;\frac{n^{res}}{n_0},
\\
&&
\label{pn_res_konacno}
\frac{p_\bot^{res}}{n_0 T_\bot} = -R_\Vert\left(\frac{T_\bot}{T_\Vert}\right)^\frac{3}{2}
\left[1 + \frac{15}{4}\frac{\zeta_1}{\varphi} + \left(\frac{15}{8}\frac{\zeta_1^2}{\varphi^2} - \frac{3}{4}\frac{\zeta_1}{\varphi}\right)\;\, g\left(\varphi/\zeta_1\right)\right]\; \varphi^\frac{3}{2},
\end{eqnarray}
where we used the notations
\begin{eqnarray}
\label{b_vert}
&& \hspace{-1cm}
R_\Vert = \frac{4}{3\sqrt\pi}\left(1-\frac{m u_z^2}{T_\Vert} - \frac{T_\Vert}{T_\Vert^{trap}}\right) \, \exp\left(-\frac{m u_z^2}{2 T_\Vert}\right),
\quad\quad 
Q_\Vert =
\frac{1}{2}\left(1-\frac{m u_z^2}{T_\Vert} - \frac{T_\Vert}{T_\Vert^{trap}}\right)^{-1},
\\
\label{notat}
&& \hspace{-1cm}
\varphi = -(q/T_\bot)(\phi-u_z A_z),
\quad\;
\zeta_1= -\delta B/B_0 - (\nabla_\bot^2/ \Omega B_0)(\phi- u_z A_z) \to -\delta B/B_0,
\quad\;
g(t)= \sqrt{\pi t}\;\exp(t)\;{\rm erfc}(\sqrt t),
\end{eqnarray}
and erfc is the complementary error function ${\rm erfc}(\sqrt t) = (2/\sqrt\pi)\int_{\sqrt t}^\infty dz\; \exp(-z^2)$. The function $g$ in Eqs. (\ref{n_res_konacno})-(\ref{pn_res_konacno}) behaves asymptotically as $g(t)\sim\tanh(\sqrt t)$, with a full agreement for both $t\to\infty$ and $t\to 0$. We note that in Eq. (\ref{notat}) we can safely set $\zeta_1 = -\delta B/B_0\equiv\zeta$, since in the regime $u_z \lesssim v_{T_\Vert}$ the term $(\nabla_\bot^2/ \Omega B_0)(\phi- u_z A_z)$ represents a small FLR correction and it is negligible in the above setting. Conversely, for $u_z > v_{T_\Vert}$, the contribution of trapped particles can be neglected altogether, since from Eq. (\ref{b_vert}) it scales as $\sim\exp(-u_z^2/v_{T_\Vert}^2)$.

\subsection{Field equations}

Now we can easily write down the Poisson's equation, and the components of the Ampere's law that are parallel and perpendicular to the magnetic field, viz.
\begin{eqnarray}
\label{poissons}
&& q_e n_e + q_i n_i = -\epsilon_0\nabla^2\phi,  \\
\label{amperes_P}
&& q_e n_e U_{e_\Vert} + q_i n_i U_{i_\Vert} = -c^2\epsilon_0 \; \vec{b}\cdot\nabla^2 \vec{A} - \epsilon_0 \; \vec{b}\cdot{\partial\vec{E}}/{\partial t} , \\
\label{amperes_N}
&& \vec{B}\times\left(q_e n_e \vec{U}_{e_\bot} + q_i n_i \vec{U}_{i_\bot}\right) = c^2\epsilon_0\left\{\left[\nabla-\vec{b}\left(\vec{b}\cdot\nabla\right)\right]{B^2}/{2}-B^2\left(\vec{b}\cdot\nabla\right)\vec{b}\right\} - \epsilon_0\; \vec{B}\times{\partial\vec{E}}/{\partial t},
\end{eqnarray}
where the densities $n_e$ and $n_i$, and the parallel velocities $U_{e_\Vert}$ and $U_{i_\Vert}$ are the sums of the respective nonresonant and resonant components for each particle species, given in Eq.s (\ref{n_nr})-(\ref{pn_nr}) and (\ref{n_res_konacno})-(\ref{pn_res_konacno}). The last terms on the right-hand-sides of Eqs. (\ref{amperes_P}) and (\ref{amperes_N}) come from the displacement current, and can be neglected for the low frequency mode we study.
The perpendicular fluid velocities $\vec{U}_{e,i_\bot}$ can not be obtained from the gyrokinetic equation, since they are related with the gyroangle-dependent component of the distribution function that is not described by Eq. (\ref{my_gyrokinetic_2}). Within the adopted accuracy, for each particle species we express them from appropriate momentum equations, viz.
\begin{equation}
\label{momentum_eq}
\left(\frac{\partial}{\partial t} + \vec{U}\cdot\nabla\right)\vec{U} = \frac{q}{m}\left(\vec{E} + \vec{U}\times\vec{B}\right) - \frac{1}{m n}\nabla\cdot\left(P + \pi\right),
\end{equation}
where, for simplicity, we have omitted the subscripts $e$ and $i$ referring to the electrons and ions, respectively. In the above, the pressure $P$ and the stress $\pi$ are diagonal and off-diagonal tensors. The temperature is assumed to be anisotropic and the pressure tensor is given by $P = p_\bot(I - \vec{b}\,\vec{b}) + p_\Vert\vec{b}\,\vec{b}$ where $I$ is a unit tensor, viz. $I_{\alpha, \beta} = \delta_{\alpha, \beta}$ and $\delta_{\alpha, \beta}$ is the Kronecker delta. We use the usual shorthand notation from vector algebra $\vec{p}\cdot\vec{q} \, \vec{r} = (\vec{p}\cdot\vec{q})\vec{r}$ and $\nabla\cdot\vec{q} \, \vec{r} = (\nabla\cdot\vec{q} + \vec{q}\cdot\nabla)\vec{r}$.
The chain of hydrodynamic equations is truncated by the use of the Braginskii's collisionless stress tensor \cite{1965RvPP....1..205B}, given below in Eq. (\ref{braginskii_G}), that are appropriate for perturbations that are weakly varying both on the timescale of the gyroperiod and on the spatial scale of the Larmor radius. Conversely, the pressures $p_\Vert$ and $p_\bot$ are given by the appropriate moments of the distribution function (\ref{free_distr}) and (\ref{trap_distr}). After multiplying the momentum equations (\ref{momentum_eq}) with $\vec{b}\times$, the perpendicular fluid velocity can be readily written as the sum of the $\vec{E}\times\vec{B}$, diamagnetic, anisotropic-temperature, stress-related [also called the FLR (finite-Larmor-radius) drift], and polarization drifts, viz.
\begin{equation}\label{drift_velocity_1}
\vec{U}_\bot = \vec{U}_E + \vec{U}_D + \vec{U}_A + \vec{U}_\pi + \vec{U}_p,
\end{equation}
where
\begin{eqnarray}
&&
\nonumber
\vec{U}_E = -\frac{\vec{b}}{B}\times\vec{E}, \quad
\vec{U}_D = \frac{\vec{b}}{q n B}\times\nabla_\bot p_\bot, \quad
\vec{U}_A = \left(p_\Vert - p_\bot\right)\frac{\vec{b}}{q n B}\times\left(\vec{b}\cdot\nabla\right)\vec{b},
\\
&&
\vec{U}_\pi = \frac{\vec{b}}{q n B}\times\vec{e}_\alpha\frac{\partial\pi_{\alpha, \beta}}{\partial x_\beta}, \quad
\vec{U}_p = \frac{\vec{b}}{\Omega}\times\left(\frac{\partial}{\partial t} + \vec{U}\cdot\nabla\right)\vec{U}, \quad
\vec{U}_B = \frac{p_\bot \vec{b}}{q n B^2}\times\nabla_\bot B,
\label{drifts}
\end{eqnarray}
and the collisionless stress tensor $\pi_{\alpha,\beta}$, under the scaling of Eqs. (\ref{gyro_drift_scaling}) and (\ref{My_scaling}), has the Braginskii's form \cite{1965RvPP....1..205B}
\begin{eqnarray}
\nonumber && \pi_{m,m} = - \pi_{l,l} = \left(p_\bot/2\Omega\right)\left(\partial U_l/\partial x_m + \partial U_m/\partial x_l\right), \\
\nonumber && \pi_{l,m} = \pi_{m,l} = \left(p_\bot/2\Omega\right)\left(\partial U_l/\partial x_l - \partial U_m/\partial x_m\right), \\
\nonumber && \pi_{l,b} = \pi_{b,l} = - \left(p_\bot/\Omega\right)\left(\partial U_m/\partial x_b + \partial U_b/\partial x_m\right), \\
\nonumber && \pi_{m,b} = \pi_{b,m} = \left(p_\bot/\Omega\right)\left(\partial U_l/\partial x_b + \partial U_b/\partial x_l\right), \\
&& \pi_{b,b} = 0,
\label{braginskii_G}
\end{eqnarray}
where we used the notation $\partial/\partial x_b = \vec{b}\cdot\nabla$ and $\vec{e}_l$, $\vec{e}_m$, and $\vec{b}$ are three mutually perpendicular unit vectors. A natural choice, for a curved magnetic field, is to adopt $\vec{e}_l$ and $\vec{e}_m$ to be parallel to the normal and to the bi-normal of the magnetic field line, respectively, viz. $\vec{e}_l = [(\vec{b}\cdot\nabla)\vec{b}]/|\nabla\times\vec{b}|$  and  $\vec{e}_m = (\nabla\times\vec{b})/|\nabla\times\vec{b}|$. As the magnetic field lines are only weakly curved, $|(\vec{b}\cdot\nabla)\vec{b}| \sim \epsilon \ll 1$, within the adopted scaling we may use instead $\vec{e}_l \approx \vec{e}_x$, $\vec{e}_m\approx \vec{e}_y$, and $\vec{b}\approx\vec{e}_z$.
We note that the stress tensor Eq. (\ref{braginskii_G}) should be used with caution, since it has been derived in the framework of the transport theory and for a collisional plasma and its reduction to a collisionless limit is not straightforward. In particular, the condition for neglecting the parallel viscosity, $\tau\ll \omega L_\Vert^2/v_T^2$ (where $\tau$ is the collision time, while $\omega^{-1}$ and $L_\Vert$ are the characteristic temporal and parallel spatial scales, respectively) was obtained assuming that the compression of magnetic field is negligible; for small-but-finite compressional perturbations, the condition for neglecting the parallel viscosity may be different.
Now, the stress-related drift velocity can be written as
\begin{eqnarray}
\label{U_pi_B}
&&
\vec{U}_\pi = \frac{1}{q n B}
\frac{p_\bot}{2 \Omega}\nabla^2_\bot\vec{U}_\bot  + \delta \vec{U}_\pi,
\\
\nonumber
&&
\delta \vec{U}_\pi = \frac{1}{q n B}\left\{
\left[\left(\vec{b}\times\nabla_\bot\frac{p_\bot}{2\Omega}\right)\cdot\nabla_\bot\right]\vec{b}\times\vec{U}_\bot + \left(\nabla_\bot\frac{p_\bot}{2\Omega}\cdot\nabla_\bot\right)\vec{U}_\bot + \right.\left. \frac{\partial}{\partial x_b}\left[\frac{p_\bot}{\Omega}\left(\frac{\partial \vec{U}_\bot}{\partial x_b} + \nabla_\bot U_\Vert\right)\right] +\frac{p_\bot}{2}\frac{\partial\vec{U}_\bot}{\partial x_b}\frac{\partial}{\partial x_b}\frac{1}{\Omega}\right\},
\\
&&
\label{delta_U_pi_B}
\end{eqnarray}
where $\nabla_\bot = \nabla - \vec{b} \; (\partial/\partial x_b)$. We note that the two terms on the right-hand-side of Eq. (\ref{U_pi_B}) scale relative to each other as $\delta^2/\epsilon$, where  $\delta = v_{T_\bot}\nabla_\bot/\Omega\ll 1$ and $\epsilon = (1/\Omega)(d/dt)\ll 1$, see Eqs. (\ref{small_param}) and (\ref{dodatno_skaliranje}). Using these small parameters, with the accuracy to leading order, we can set $\vec{U}_\bot \approx \vec{U}_\bot^{apr}$, where the approximative fluid velocity $\vec{U}_\bot^{apr}$ is determined from
\begin{equation}\label{U_priblizno_B}
\left(1 - \rho_{L_0}^2\nabla^2_\bot/2\right)\vec{U}_\bot^{apr} = \vec{U}_E + \vec{U}_D,
\end{equation}
and $\rho_{L_0} = \sqrt{T_\bot/m\Omega_0^2}$ is the (unperturbed) Larmor radius. The divergence of the perpendicular component of the Ampere's law (\ref{amperes_N}) takes the simple form of a pressure balance equation
\begin{equation}\label{lead_pres_bal}
\nabla \cdot\left[\nabla_\bot\left({c^2\epsilon_0 B^2}/{2} + p_{i_\bot} + p_{e_\bot}\right)\right] =  \delta{F}_1 + \delta{F}_2 + \delta{F}_3 + \delta{F}_4,
\end{equation}
where the small corrections on the right-hand-side are given by
\begin{eqnarray}
\label{delta_F_1}
&&
\delta{F}_1 = \nabla\cdot\left[\left(c^2\epsilon_0 B^2 + p_{i_\bot} - p_{i_\Vert} + p_{e_\bot}- p_{e_\Vert}\right) \; \frac{\partial\vec{b}}{\partial x_b}\right],
\\
\label{delta_F_2}
&&
\delta{F}_2 = - \sum_{j=e,i} \nabla\cdot\left[\vec{b}\times \frac{p_{j_\bot}}{2\Omega_j} \, \nabla_\bot^2 \vec{U}_{j_\bot}^{apr}\right] ,
\\
\label{delta_F_3}
&&
\delta{F}_3 = \sum_{j=e,i} \nabla\cdot\left[q_j n_j \vec{B}\times \delta\vec{U}_{j \pi} -m_j n_j \left(\frac{\partial}{\partial t} + \vec{U}_{j_\bot}^{apr}\cdot\nabla\right)\vec{U}_{j_\bot}^{apr}\right],
\\
\label{delta_F_4}
&&
\delta{F}_4 = \epsilon_0 \, \nabla\cdot\left[\vec{B}\times\frac{\partial\vec{E}}{\partial t} + \vec{E}_\bot\left(\nabla\cdot\vec{E}\right)\right].
\end{eqnarray}
These terms arise due to the curvature of magnetic field lines, FLR effects, particle polarization drifts, and due to displacement current and charge separation, respectively. Their scalings relative to the left-hand-side of Eq. (\ref{lead_pres_bal}) are given by $(\partial/\partial x_b)/\nabla_\bot$, $\delta^2$, $\epsilon$, and $(d^2/dt^2)/(c^2 \nabla_\bot^2)$, respectively, and thus they all can be neglected within the orderings of Eqs. (\ref{small_param}) and (\ref{dodatno_skaliranje}). The small contribution of the displacement current in the parallel component of the Amperes's law, i.e. the last term on the right-hand-side of Eq. (\ref{amperes_P}), will be neglected, too.
As it is evident from its derivation, the pressure balance equation in the form (\ref{lead_pres_bal}) is valid only if the parallel convective derivative can be neglected, $U_\Vert (\partial/\partial x_b)\to 0$, or equivalently $U_\Vert\ll u_z$. Obviously, this holds in the linear regime, but also in the nonlinear regime, provided $u_z^2>\epsilon \, v_{T_\Vert}^2$, see Eq. (\ref{Up_nr}).

It is convenient to represent the electromagnetic field in terms of three scalar quantities that are the electrostatic potential $\phi$ and the components of the magnetic field and of the vector potential parallel to the unperturbed magnetic field, $B_z$ and $A_z$, respectively. Then the normal (to $z$) components of these vectors are found from
\begin{equation}\label{perpAB}
\nabla_{\! n}^2 \, \vec{B}_n = -\vec{e}_z\times\nabla \, \nabla^2 A_z - \nabla_{\! n} \left({\partial B_z}/{\partial z}\right)
\quad\quad {\rm and} \quad\quad
\nabla_{\! n}^2 \, \vec{A}_n = \vec{e}_z\times\nabla B_z - \nabla_{\! n} \left({\partial A_z}/{\partial z}\right),
\end{equation}
where we used the notation $\delta B_z = B_z - B_0$ and the subscript $n$ denotes the vector component normal to $z$ axis, $\vec{\xi}_n = \vec{e}_x \, \xi_x + \vec{e}_y \, \xi_y$.
To leading order in $\epsilon$, we can now write
$
\nabla_{\! n}\cdot(\partial\vec{b}/\partial x_b) \approx  - \left({1}/{B_0}\right)({\partial^2 B_z}/{\partial z^2}),
$ 
which permits us to rewrite the pressure balance equation (\ref{lead_pres_bal}) a compact form, viz.
\begin{eqnarray}\nonumber
&&
\!\!\!\!\!\!
{\nabla_n^2}\left[c^2\epsilon_0 B_0^2 \; \frac{\delta B_z}{B_0} + \delta p_{i_\bot} + \delta p_{e_\bot} +
\frac{\rho_{Li}^2}{2}\;\nabla^2\left(q_i n_0 \phi + \delta p_{i_\bot}\right)\right] =
\\
&&
\!\!\!\!\!\!
-\left[c^2\epsilon_0 B_0^2 + p_{0 i_\bot} - p_{0 i_\Vert} + p_{0 e_\bot}- p_{0 e_\Vert}- u_z^2 \; m_i n_0  \left(1 + \frac{\Omega_i^2}{\omega_{p,i}^2}\right) \right] \frac{\partial^2}{\partial z^2} \frac{\delta B_z}{B_0} ,
\label{noconv_pres_bal}
\end{eqnarray}
In the above we used $\partial/\partial t = -u_z \; \partial/\partial z$ and we retained only the leading FLR and polarization corrections related with ions and dropped those related with electrons, due to their small mass, $m_e/m_i\ll 1$. Note also that in the solar wind region of our interest we have $c_A = c \Omega_i/\omega_{p, i}\sim c_s\ll c$, which implies $\Omega_i/\omega_{p, i}\ll 1$.

Appropriately adjusting Eqs. (\ref{n_nr})-(\ref{pn_nr}) by the substitution $\nabla_\bot^2\phi\to \nabla_\bot^2(\phi + \delta p_\bot/q n_0)$, where the leading-order expression $\delta p_\bot/n_0 T_\bot\approx -Z_R \,\; (q/T_\Vert)(\phi-u_z A_z) + 2(1-Z_R \,\; T_\bot/T_\Vert)(\delta B/B_0)$ is employed, and using Eq. (\ref{noconv_pres_bal}), we obtain the following coupled equations for the parallel magnetic field and the electrostatic and vector potentials
\begin{eqnarray}
\nonumber
&&\hspace{-.5cm}
\left(Z_{R i} \, \frac{T_0}{T_{i_\Vert}} + Z_{R e} \, \frac{T_0}{T_{e_\Vert}}\right)\,\varphi + \left(Z_{R i} \, \frac{T_{i_\bot}}{T_{i_\Vert}} - Z_{R e} \, \frac{T_{e_\bot}}{T_{e_\Vert}}\right) \, \zeta -
\rho_0^2\nabla_\bot^2
\left[\left(1- Z_{R i} \, \frac{T_{i_\bot}}{T_{i_\Vert}}\right)^2\left(\varphi + 2\zeta\, \frac{T_{i_\bot}}{T_0}\right) + \psi\right] =
\frac{n_e^{res}-n_i^{res}}{n_0},
\\
\label{Poison_fin}
\\
&&\hspace{-.5cm}
\nonumber
\rho_0^2\nabla_\bot^2
\left\{\left(1- Z_{R i} \, \frac{T_{i_\bot}}{T_{i_\Vert}}\right)\left(\varphi + 2\zeta\, \frac{T_{i_\bot}}{T_0}\right) - \frac{T_0}{m_i u_z^2} \left[\frac{c_A^2 - u_z^2}{T_0/m_i} + \frac{T_{i_\bot}}{T_0}\left(1-\frac{T_{i_\Vert}}{T_{i_\bot}}\right) + \frac{T_{e_\bot}}{T_0}\left(1-\frac{T_{e_\Vert}}{T_{e_\bot}}\right)\right]\psi\right\} =
\\
&&
\label{Ampere_fin}
\frac{\left(n_e U_{e_\Vert}\right)^{res} - \left(n_i U_{i_\Vert}\right)^{res}}{n_0 u_z} - \frac{n_e^{res}-n_i^{res}}{n_0},
\end{eqnarray}
\begin{eqnarray}\nonumber
&&\hspace{-.5cm}
\nabla_\bot^2\left\{
\left(Z_{R i} \, \frac{T_{i_\bot}}{T_{i_\Vert}} - Z_{R e} \, \frac{T_{e_\bot}}{T_{e_\Vert}}\right) \, \varphi -
\left[\frac{c_A^2}{T_0/m_i} + 2\frac{T_{i_\bot}}{T_0}\left(1-Z_{R i} \,\frac{T_{i_\Vert}}{T_{i_\bot}}\right) + 2\frac{T_{e_\bot}}{T_0}\left(1-Z_{R e} \,\frac{T_{e_\Vert}}{T_{e_\bot}}\right)\right] \, \zeta + \frac{p_{e_\bot}^{res} + p_{i_\bot}^{res}}{n_0 T_0} -
\right.
\\
&&\hspace{-.5cm}
\nonumber
\left.
\rho_0^2\nabla_\bot^2 \; \frac{5}{2} \frac{T_{i_\bot}}{T_0}
\left[\left(1- Z_{R i} \, \frac{T_{i_\bot}}{T_{i_\Vert}}\right)\left(1- Z_{R i} \, \frac{4 \, T_{i_\bot}}{5 \, T_{i_\Vert}}\right)\left(\varphi + 2\zeta\, \frac{T_{i_\bot}}{T_0}\right) + \psi\right]\right\} =
\\
&&\label{pres_bal_fin}
\left[\frac{c_A^2 - u_z^2}{T_0/m_i} + \frac{T_{i_\bot}}{T_0}\left(1-\frac{T_{i_\Vert}}{T_{i_\bot}}\right) + \frac{T_{e_\bot}}{T_0}\left(1-\frac{T_{e_\Vert}}{T_{e_\bot}}\right)\right]
\frac{\partial^2\zeta}{\partial z^2}
\end{eqnarray}
where the nonlinear terms coming from the contributions of resonant particles are given by Eqs. (\ref{n_res_konacno})-(\ref{pn_res_konacno}) and
where we used the notations $\varphi = -(e/T_0)(\phi - u_z A_z)$, $\psi = -(e/T_0) \; u_z A_z$, $\zeta = -\delta B/B_0$, $\rho_0 = \sqrt{T_0/m_i \Omega_i^2}$, and $T_0$ is a normalization constant that will be conveniently adopted later.

\subsection{Linear dispersion relation and coherent nonlinear solutions}

Equations (\ref{Poison_fin})-(\ref{pres_bal_fin}) can be decoupled after some straightforward, albeit tedious algebra. Within the adopted scaling, we may neglect the higher-order FLR terms and the FLR corrections in the nonlinear terms arising from the trapped particles' contributions, which implies that we may set also $\zeta_1 \to \zeta,$ in Eq. (\ref{notat}). Thus, we arrive at
\begin{equation}\label{eq_for_zeta}
\nabla_\bot^2\left[\left(D_\bot\nabla_\bot^2 - L\right)\zeta + S\, \zeta^\frac{3}{2}\right] - D_\Vert \, \frac{\partial^2\zeta}{\partial z^2} = 0.
\end{equation}
From Eq. (\ref{eq_for_zeta}) we obtain the well-known linear dispersion relation $D_\bot \, k_\bot^2 + L + D_\Vert \, k_z^2/k_\bot^2 = 0$. However, for a complete linear picture (necessary e.g. to obtain the growthrate of the magnetic ion mirror mode), instead of the trapped particles' contributions (\ref{n_res_konacno})-(\ref{pp_res_konacno}) we need to include also the small linear contribution of resonant particles via the Landau damping, by using the complete plasma dispersion function $Z_\Vert$ rather than its real part $Z_R$ in Eqs. (\ref{n_nr})-(\ref{pn_nr}). The constant coefficients $D_\Vert$, $D_\bot$, $L$, and $S$ depend in a complicated way on the plasma parameters $n_{e,i}$, $T_{e,i_\bot}$, $T_{e,i_\Vert}$, and also on the parallel phase velocity $u_z$ and on the effective temperatures of trapped particles, $T^{trap}_{e,i_\Vert}$. Below, we will discuss in more details only the special cases of isothermal and adiabatic ions, $u_z\ll v_{Ti\Vert}$ and $u_z\gg v_{Ti\Vert}$, in which the nonlinear dynamics is governed by the trapped ions and trapped electrons, respectively.

Analytically, we can find two distinct coherent nonlinear solutions of the above equation for the compressional magnetic field. In the simple one-dimensional case when  $\zeta$ depends only on $z_1 = y/\theta + z$, where $\theta={\rm constant}\ll 1$, we have $\nabla_\bot^2 = \partial^2/\partial y^2 = \theta^{-2}\;\partial^2/\partial z^2$ and, using normalized variables 
$z' =  z\;\theta\sqrt{(L+D_\Vert\theta^2)/D_\bot}$ and $\zeta' = \zeta \; S^2/(L+D_\Vert\theta^2)^2$,
our Eq. (\ref{eq_for_zeta}) simplifies to
\begin{equation}\label{eq_for_zeta_31}
\left(d^2/dz^{\prime \, 2}\right)\left[d^2\zeta'/dz^{\prime \, 2}-\zeta' + {\rm sign}\left(S\right) \; \zeta^{\prime \, \frac{3}{2}}\right] = 0.
\end{equation}
When both $S>0$ and $(L+D_\Vert\theta^2)/D_\bot>0$, this nonlinear equation readily yields a localized solution in the form
\begin{equation}\label{resenje_slab}
\zeta'\left(z'\right) = \left(25/16\right) \; \cosh^{-4}\left(z'/4\right),
\end{equation}
if the boundary condition $\zeta'(0)=25/16$ is used. Conversely, if $\zeta'(0)$ is adopted to be somewhat smaller than $25/16$, a periodic structure (cnoidal wave) is obtained instead of the solitary solution.

Alternatively, a spheroidal solution can be found when the dependence along the magnetic field is sufficiently weak, $D_\Vert \; \partial^2/\partial z^2\ll L \; \nabla_\bot^2$, so that Eq. (\ref{eq_for_zeta}) can be rewritten as
\begin{equation}\label{eq_for_zeta_2}
\left(\nabla_\bot^2 + \frac{D_\Vert}{L}\frac{\partial^2}{\partial z^2}\right)\left[D_\bot\left(\nabla_\bot^2 - \frac{D_\Vert}{L}\frac{\partial^2}{\partial z^2}\right)\zeta - L \; \zeta + S\, \zeta^\frac{3}{2}\right] = - \frac{D_\Vert}{L}\frac{\partial^2}{\partial z^2}\left(D_\bot \; \frac{D_\Vert}{L}\frac{\partial^2\zeta}{\partial z^2} - S\, \zeta^\frac{3}{2}\right) \to 0.
\end{equation}
For a 1-D solution that depends only on the quantity $r = \sqrt{x^2+y^2-z^2\;L/D_\Vert}$, introducing normalized variables $r'= r\sqrt{L/D_\bot}$ and  $\zeta' = \zeta \; S^2/L^2$, our Eq. (\ref{eq_for_zeta_2}) is rewritten in a simple form, viz.
\begin{equation}\label{eq_for_zeta_3}
\nabla^{\prime \, 2}\,\zeta'-\zeta'+ {\rm sign}\left(S\right) \; \zeta^{\prime \, \frac{3}{2}} = 0,
\quad\quad
\nabla^{\prime \, 2}=\partial^2/\partial r^{\prime \, 2} + (2/r')(\partial/\partial r').
\end{equation}
This equation is easily integrated, yielding (for $L/D_\bot>0$) a bell-shaped localized solution similar to that given by Eq. (\ref{resenje_slab}), albeit with a bigger maximum. When $L/D_\Vert<0$, this corresponds to a cigar-shaped structure in three spatial dimensions, elongated along the magnetic field. Conversely, when $L/D_\Vert>0$ it can be visualized as a one-sheeted hyperboloid whose axis of symmetry is parallel to the magnetic field, comprising a topological x-point in space. Such solution becomes singular when $z^2 \geq (D_\Vert/L)(x^2+y^2)$,
and we may speculate that the exact solution, that contains higher powers of $\partial/\partial z$ and permits larger values of $\zeta'$, might have the form of a periodic chain of magnetic depressions separated by such x-points, whose wavelength along the magnetic field is of the order $\lambda_z \sim \pi\sqrt{D_\Vert D_\bot/L^2}$.
{In principle, other shapes of the structure might also be possible, e.g. an infinitely long cylinder oblique to the magnetic field, etc. However, for those one cannot reduce our equation (\ref{eq_for_zeta}) to an essentially 1-D geometry (slab, spherical, cylindrical) and the solution would require extensive 3-d numerical calculations that are out of scope of the present paper.}
\begin{figure}
\includegraphics[width=80mm]{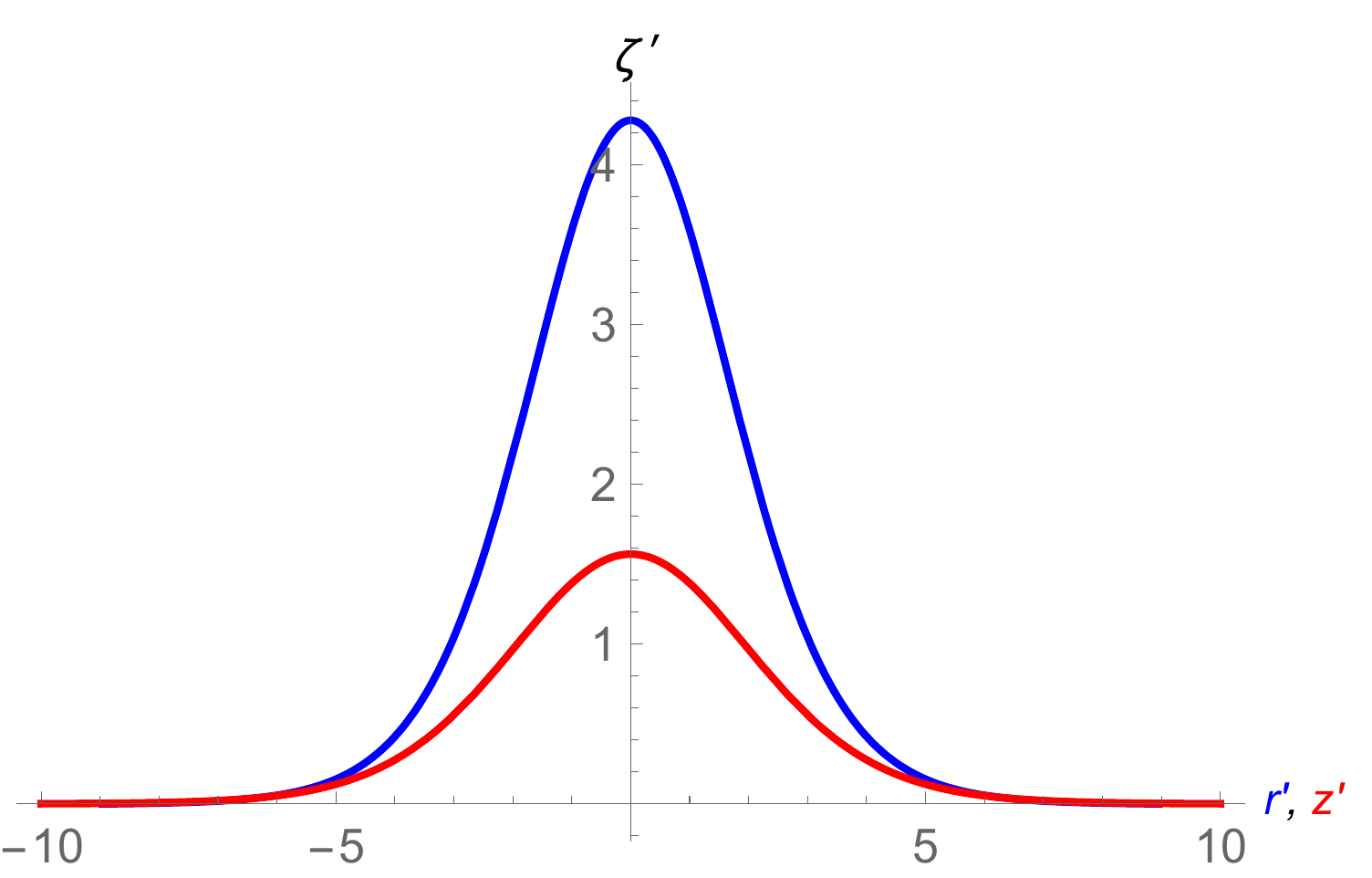}
\caption{Oblique slab (red) and spheroidal (blue) magnetic holes $\zeta'\propto -\delta B_z/B_0$, determined from Eq. (\ref{eq_for_zeta}). (color online) } \label{slab_spheroid}
\end{figure}

\subsubsection{Magnetic ion mirror mode}

The simple theory of the magnetic ion mirror mode \cite{16,19,Kuznetsov_PhRvL} is applicable in plasmas in which the parallel temperature of ions is much bigger than that of electrons, viz. $T_{e_\Vert}/T_{i_\Vert}\ll m_e/m_i$. Then, for the perturbations whose parallel phase velocity satisfies $v_{T e_\Vert}\ll u_z\ll v_{T i_\Vert}$, the real parts of the plasma dispersion functions of electrons and ions are given by $Z_{R e}=-v_{T e_\Vert}^2/u_z^2$ and $Z_{R i}= 1$, respectively. We note that the contributions of trapped electrons are negligibly small $\propto\exp(-u_z^2/v_{Te_\Vert}^2)$ and conveniently adopt the normalization in Eqs (\ref{Poison_fin})-(\ref{pres_bal_fin}) to be $T_0 = T_{i_\bot}$. In the given range of parallel phase velocities, we readily find from the Poisson's equation (\ref{Poison_fin}) that the parallel electric field is small (i.e. of the order of FLR corrections), $\varphi\sim\rho_{Li}^2\nabla_\bot^2\zeta$, while the Ampere's law (\ref{Ampere_fin}) yields that the vector potential scales as $\psi\sim\zeta\;(u_z^2/c_A^2)$, which is negligibly small in the plasmas of our interest, in which the Alfv\`{e}n speed is of the same order as the parallel ion thermal velocity $c_A\sim v_{Ti_\Vert}\gg u_z$. Thus, after some straightforward algebra, the pressure balance equation (\ref{pres_bal_fin}) can be cast in the same form as the nonlinear equation (\ref{eq_for_zeta}), with the following values of the coefficients
\begin{eqnarray}
\nonumber
&&
D_\Vert={c_A^2}/{v_{T i_\bot}^2}+1-{T_{i_\Vert}}/{T_{i_\bot}},
\quad\quad\;\;\;
D_\bot = -\rho_{Li}^2\left(1-{T_{i_\bot}}/{T_{i_\Vert}}\right)\left(5-4\;{T_{i_\bot}}/{T_{i_\Vert}}\right),
\\
\label{coef_mirror}&&
L={c_A^2}/{v_{T i_\bot}^2}+2\left(1-{T_{i_\bot}}/{T_{i_\Vert}}\right),
\quad\quad
S=-\left({4}/{3\sqrt\pi}\right)\left(1-T_{i_\Vert}/T_{i_\Vert}^{trap}\right)\left(T_{i_\bot}/T_{i_\Vert}\right)^\frac{3}{2},
\end{eqnarray}
which implies that localized solutions (\ref{resenje_slab}) and (\ref{eq_for_zeta_3}) exist if both the trapped ions create a hump in the distribution function, $0<T_{i_\Vert}^{trap}<T_{i_\Vert}$ and the ion temperature anisotropy is in the range $1<T_{i_\bot}/T_{i_\Vert}<{\rm min}(5/4, \; c_A^2/2 v_{T i_\bot}^2)$.

We note from Eqs. (\ref{Poison_fin})-(\ref{pres_bal_fin}) that in the case of a finite electron temperatute, $T_{e_\Vert}\ne 0$, the parallel electric field remains finite and the expressions for the coefficients $D_{\Vert, \bot}$, $L$ and $S$ become rather complicated. The warm electron effects were studied in \cite{2009PhPl...16l2901I}. They were shown to reduce the growth rate of the mirror instability, as the electrons are dragged by nonresonant ions that are mirror accelerated from regions of high to low parallel magnetic flux. The mirror mode's nonlinear dynamics is also affected, so that the transition from the linear to nonlinear regime occurred when the wave amplitude was $\sim50\%$ of that required in the cold electron temperature limit. In the further nonlinear dynamics, the explosive formation of magnetic holes took place and saturated into cnoidal waves or solitary structures, and it was shown that the finite electron temperature decreases the holes' spatial dimensions and increases their depth.

\subsubsection{Field swelling and kinetic Alfv\'{e}n modes}

In the regime of parallel phase velocities that lie between the parallel electron and ion thermal speeds, $v_{T i_\bot}\ll u_z \ll v_{T e_\bot}$, setting $Z_{R e}=1$ and $Z_{R i}= -v_{T i_\Vert}^2/u_z^2$, noting that the contributions of trapped ions are negligibly small $\propto\exp(-u_z^2/v_{Ti_\Vert}^2)$, and adopting the normalization $T_0 = T_{e_\bot}$, our basic equations Eqs. (\ref{Poison_fin})-(\ref{pres_bal_fin}) are simplified to
\begin{eqnarray}
&&\hspace{-.5cm}
\left(\frac{T_{e_\bot}}{T_{e_\Vert}}-\frac{c^2_{S_\bot}}{u_z^2}\right)\,\varphi - \frac{T_{e_\bot}}{T_{e_\Vert}} \, \zeta -
\rho_{S_\bot}^2\nabla_\bot^2 \left(\varphi + 2\, \frac{T_{i_\bot}}{T_{e_\bot}}\,\zeta + \psi \right) =
\frac{n_e^{res}}{n_0},
\label{Poisson_EH}
\\
&&\hspace{-.5cm}
\left(\frac{T_{e_\bot}}{T_{e_\Vert}}-\frac{c^2_{S_\bot}}{u_z^2}\right)\,\varphi - \frac{T_{e_\bot}}{T_{e_\Vert}} \, \zeta -
{\cal D} \; \rho_{S_\bot}^2\nabla_\bot^2 \;\psi = \frac{\left(n_e U_{e_\Vert}\right)^{res}}{n_0 u_z},
\label{Ampere_EH}
\\
&&\hspace{-.5cm}
\nonumber
\nabla_\bot^2\left[\left(\frac{T_{e_\bot}}{T_{e_\Vert}}+\frac{v_{T i_\bot}^2}{u_z^2}\right)\, \varphi +
2 \left(\frac{c_A^2}{2 c_{S_\bot}^2} + \frac{T_{i_\bot}}{T_{e_\bot}} + 1-\frac{T_{e_\Vert}}{T_{e_\bot}}\right) \,\zeta +
\frac{5}{2}\frac{T_{i_\bot}}{T_{e_\bot}} \; \rho_{S_\bot}^2\nabla_\bot^2 \left(\varphi + 2\, \frac{T_{i_\bot}}{T_{e_\bot}}\,\zeta + \psi\right) \right] -
\\&&\hspace{-.5cm}
\label{pres_bal_EH}
\left(1-{\cal D}\right)\frac{u_z^2}{c_{S_\bot}^2}\frac{\partial^2\zeta}{\partial z^2}  =
\nabla_\bot^2 \, \frac{p_{e_\bot}^{res}}{n_0 T_{e_\bot}},
\end{eqnarray}
where $c_{S_\bot}=\sqrt{T_{e_\bot}/m_i}$, $\rho_{S_\bot}=c_{S_\bot}/\Omega_i$, and  
${\cal D} = ({c_{S_\bot}^2}/{u_z^2})
[{c_A^2}/{c_{S_\bot}^2} + ({T_{i_\bot}}/{T_{e_\bot}})(1-{T_{i_\Vert}}/{T_{i_\bot}}) + 1-{T_{e_\Vert}}/{T_{e_\bot}} ]$.
In a plasma with sufficiently cold electrons, $\beta_{e_\bot}\ll 1$, where $\beta_{e_\bot} = 2 c_{S_\bot}^2/c_A^2 = 2 n_0 T_{e_\bot}/c^2\epsilon_0 B^2$, and for the parallel phase velocities satisfying $c_A\gg u_z \gg c_{S_\bot}$ the above reduce to the 
equation for nonlinear electrostatic drift waves \cite{2000PhRvL..84.4373J}, $(1-\rho_{S_\bot}^2 \nabla_\bot^2)\varphi = n_e^{res}/n_0$, that feature no magnetic field  perturbations, $\psi=\zeta=0$, and possess the 
solutions displayed in Fig. \ref{slab_spheroid}.
Conversely, in a hot electron plasma, $\beta_{e_\bot} = 2 c_{S_\bot}^2/c_A^2 \gtrsim 1$, phase velocities in the range $v_{T i_\bot}<u_z<v_{T e_\bot}$ correspond to the magnetosonic modes which can be interpreted as linearly coupled kinetic Alfv\'{e}n and acoustic waves, that can be unstable under certain conditions  \cite{1982PhRvL..48..799B}. The unstable fast magnetosonic wave is often called the field swelling mode, while the instability of the slow magnetosonic mode is referred to as the electron mirror instability \cite{17}. The short wavelength fast magnetosonic wave, with $\rho_{L_i}\nabla_\bot\gtrsim 1$, is usually called the kinetic Alfv\'{e}n wave and it is destabilized by the inhomogeneities of the density and magnetic field \cite{2005ChA&A..29....1D} and by the electron temperature anisotropy \cite{2010PhPl...17f2107C}.

In the kinetic Alfv\'{e}n or fast magnetosonic regime, phase velocity exceeds the acoustic and Alfv\'{e}n speeds, $u_z\gg {\rm max}\,(c_{S_\bot}, c_A)$, and using the smallness of ${\cal D}$ and of the nonlinear terms, we express
$\varphi$ and $\psi$ from Eqs. (\ref{Poisson_EH}), (\ref{Ampere_EH}) as
\begin{equation}
\varphi = \left[1- {\cal D} 
\left(1+2 
\frac{T_{i_\bot}}{T_{e_\bot}}\right)\rho_{S_{\bot}}^2\nabla_\bot^2\right]\zeta + \frac{T_{e_\Vert}}{T_{e_\bot}}\frac{\left(n_e U_\Vert\right)^{res}}{n_0 u_z},
\quad\quad
\rho_{S_{\bot}}^2\nabla_\bot^2\psi = -\left(1+2 
\frac{T_{i_\bot}}{T_{e_\bot}}\right)\rho_{S_{\bot}}^2\nabla_\bot^2\zeta - \frac{n_e^{res}}{n_0},
\label{psi_fast_w}
\end{equation}
which after the substitution into Eq. (\ref{pres_bal_EH}) yields an equation for the compressional magnetic field perturbation, having the form of the generic equation (\ref{eq_for_zeta}), $\nabla_\bot^2[(D_\bot\nabla_\bot^2 - L)\zeta + S\; \zeta^\frac{3}{2}] - D_\Vert \; {\partial^2\zeta}/{\partial z^2} = 0$, with the following coefficients
\begin{eqnarray}
&& \nonumber
\hspace{-.3cm}
D_\Vert 
= -\frac{u_z^2}{c_{S_\bot}^2},
\quad\quad\quad
D_\bot = {\cal D} \; \rho_{S_\bot}^2 \; 
\frac{T_{e_\bot}}{T_{e_\Vert}}\left(1+\frac{5}{2} 
\frac{T_{i_\bot}}{T_{e_\bot}}\right)\left(1+2\; 
\frac{T_{i_\bot}}{T_{e_\bot}}\right),
\quad\quad\quad
L = \frac{T_{e_\bot}}{T_{e_\Vert}} + 2\left(\frac{c_A^2}{2 c_{S_\bot}^2} + \frac{T_{i_\bot}}{T_{e_\bot}} + 1 - \frac{T_{e_\Vert}}{T_{e_\bot}}\right)
\\
&&
\nonumber
\hspace{-.3cm}
S \equiv \zeta^{-\frac{3}{2}}\left[\frac{p_{e_\bot}^{res}}{n_0 T_{e_\bot}} + \frac{5}{2}\frac{T_{i_\bot}}{T_{e_\bot}}\frac{n_e^{res}}{n_0} - \left(1 + \frac{5}{2}\frac{T_{i_\bot}}{T_{e_\bot}}\right)\frac{\left(n_e U_{e_\Vert}\right)^{res}}{n_0 u_z}\right] =
\\
&&
\hspace{-.3cm} \hspace{.6cm}
-\frac{4}{3\sqrt{\pi}}\left\{\left(\frac{9}{8}+\frac{3 g_1}{8}\right)\left(1-\frac{T_{e_\Vert}}{T_{e_\Vert}^{trap}}\right)-\left(\frac{5}{4}+\frac{3 g_1}{8}\right)\left(1+\frac{5}{2}\frac{T_{i_\bot}}{T_{e_\bot}}\right)\right\},
\label{coeff_electr_mirror}
\end{eqnarray}
where $g_1 \equiv g(1) = \sqrt{\pi}\;e\;{\rm erfc}(1) \approx 0.7578$. We note from the above that $D_\Vert<0$ and $D_\bot>0$, and that in a plasma with anisotropic temperature $T_{e_\bot}+T_{i_\bot}> T_{e_\Vert}$ we also have $L>0$. Thus, the localized solutions associated with the nonlinear magnetosonic mode displayed in Fig. \ref{slab_spheroid} exist when $S>0$, that is fulfilled for a sufficiently small positive temperature of trapped electrons, i.e. when the trapped electrons form a small hump un the distribution function.

\subsubsection{Electron mirror mode}
{When  the ion temperature is sufficiently small, $T_{i_\bot}\ll T_{e_\bot}$, our  Eqs. (\ref{Poisson_EH})-(\ref{pres_bal_EH}) describe also the slow magnetosonic mode \cite{1982PhRvL..48..799B,17} in a high-$\beta_{e_\bot}$ plasma with $u_z\sim c_{S_\bot}\gg v_{T i_\bot}$. Such regime is not of interest for the present study of the solar wind plasma, in which the electron and ion temperatures  are of the same order and for conciseness we do not show here the corresponding lengthy coefficients $D_{\Vert}, D_\bot, L$, and $S$. We give here only the result for nonlinear slow magnetosonic structures in a plasma with warm ions, $T_{i_\bot}\sim T_{e_\bot}$, whose parallel phase velocity is smaller than the ion thermal speed, $u_z\ll v_{T i_\Vert} \sim c_{S_\bot}\ll v_{T e_\Vert}$. Using Eqs. (\ref{Poison_fin})-(\ref{pres_bal_fin}) and after some algebra, we obtain
\begin{eqnarray}
&&  \hspace{-.5cm} \label{DPSMS}
D_\Vert = -\left[\frac{c_A^2}{c_{S_\bot}^2}+\frac{T_{i_\bot}}{T_{e_\bot}}\left(1-\frac{T_{i_\Vert}}{T_{i_\bot}}\right) + 1-\frac{T_{e_\Vert}}{T_{e_\bot}}\right],
\\
&&  \hspace{-.5cm} \label{DNSMS}
D_\bot= \rho_{S_\bot}^2 \, \left(1-\frac{T_{i_\bot}}{T_{i_\Vert}}\right)^2
\left[\frac{T_{i_\bot}/T_{i_\Vert}-T_{e_\bot}/T_{e_\Vert}}{T_{e_\bot}/T_{i_\Vert}+T_{e_\bot}/T_{e_\Vert}} - \frac{5}{2}\frac{T_{i_\bot}}{T_{e_\bot}} \frac{1-(4/5)\left(T_{i_\bot}/T_{i_\Vert}\right)}{1-T_{i_\bot}/T_{i_\Vert}}\right]
\left(\frac{T_{i_\bot}/T_{i_\Vert}-T_{e_\bot}/T_{e_\Vert}}{T_{e_\bot}/T_{i_\Vert}+T_{e_\bot}/T_{e_\Vert}} - 2\frac{T_{i_\bot}}{T_{e_\bot}}\right),
\\
&&  \hspace{-.5cm} \label{LSMS}
L=-\left[\frac{\left(T_{i_\bot}/T_{i_\Vert}-T_{e_\bot}/T_{e_\Vert}\right)^2}{T_{e_\bot}/T_{i_\Vert}+T_{e_\bot}/T_{e_\Vert}} + \frac{c_A^2}{c_{S_\bot}^2}+2\frac{T_{i_\bot}}{T_{e_\bot}}\left(1-\frac{T_{i_\Vert}}{T_{i_\bot}}\right) + 2\left( 1-\frac{T_{e_\Vert}}{T_{e_\bot}}\right)\right],
\\
&&  \hspace{-.5cm} \label{SSMS}
S \equiv \zeta^{-\frac{3}{2}}\left(\frac{p_{e_\bot}^{res} + p_{i_\bot}^{res}}{n_0 \; T_{e_\bot}} + \frac{T_{i_\bot}/T_{i_\Vert}-T_{e_\bot}/T_{e_\Vert}}{T_{e_\bot}/T_{i_\Vert}+T_{e_\bot}/T_{e_\Vert}} \; \frac{n_e^{res}-n_i^{res}}{n_0}\right).
\end{eqnarray}
We note that, due to their small velocity, the slow magnetosonic structures may trap both electrons and ions which increases the number of free parameters of the problem. From the above equations we may, in principle, determine the domain in the space of plasma parameters in which the slow magnetosonic structures may exist. However, this is a very tedious task due to the complexity of Eqs. (\ref{DPSMS})-(\ref{SSMS}) and the large number of free parameters involved. }

\section{Concluding remarks}

We have studied the nonlinear effects of the trapping of resonant particles, by the combined action of the electric field and the magnetic mirror force, on different branches of the mirror and magnetoacoustic modes. In the regime of small but finite perturbations of the magnetic field, that possesses both the compressional and torsional components, we used the gyrokinetic description modified so as to include the finite Larmor radius effects that give rise to the spatial dispersion. We demonstrated that in the magnetosheath plasma, featuring cold electrons and warm ions with an anisotropic temperature, coherent nonlinear magnetic depression may be created that are associated with the nonlinear mirror mode and supported by the population of trapped ions forming a hump in the distribution function. These objects may appear either isolated or in the form of a train of weakly correlated structures (the cnoidal wave). Coherent magnetic holes of the same form appear also in the Solar wind and in the magnetopause, characterized with anisotropic electron and ion temperatures that are of the same order of magnitude. These magnetic holes are attributed to the two branches of the nonlinear magnetosonic mode, the electron mirror and the field swelling mode, including also the kinetic Alfv\'{e}n mode, supported by the population of resonant electrons that are trapped and create a small hump in the distribution function. The ion mirror, the field swelling and the electron mirror modes are described by the same generic nonlinear equation, which possesses localized solutions in the form of an oblique slab or of an ellipsoid parallel to the magnetic field and strongly elongated along it. It is worth noting that the oblique slabs appear as moving normally to the magnetic field with the velocity $u_z \theta$ (where $\theta$ is the small angle between the slab and the magnetic field), while the spheroidal "cigars" propagate strictly along $\vec B$ and, possibly, are convected in the perpendicular direction by a plasma flow. The transverse spatial scale of the ion mirror and magnetosonic structures is governed by the finite ion Larmor radius effects and may exceed $\rho_{Li}$ several times. While the ion mirror structures are purely compressional magnetic, with a negligible electric field and magnetic torsion, the magnetosonic and kinetic Alfv\'{e}n structures feature both the magnetic torsion and the finite electrostatic potential, but the ratio of the perpendicular and parallel magnetic fields is small and can be estimated from Eq. (\ref{psi_fast_w}) as $\delta B_\bot/\delta B_\Vert \sim (c_{S_\bot}/u_z) (\rho_{S_\bot}|\nabla_\bot\psi|/\zeta) \ll 1$. Our results provide a theoretical explanation for the kinetic Alfv\'{e}n magnetic holes recently observed by the NASA's Magnetospheric Multiscale (MMS) mission \cite{2017NatCo...814719G} in the Earth's magnetopause. The distribution function within those structures clearly exhibited loss-cone features, since their magnetic mirror force was sufficient to trap electrons propagating with the magnetic pitch angles between $75^{\rm o}$ and $105^{\rm o}$.

\begin{acknowledgements} This work was supported in part (D.J. and M.B.) by the MPNTR 171006 
and NPRP 8-028-1-001 grants. D.J. acknowledges financial support from the Observatoire de Paris and of the French CNRS, and the hospitality of the LESIA laboratory in Meudon.
\end{acknowledgements}

\bibliography{Meudon_2014}

\end{document}